\documentclass[modern]{aastex631}

\shorttitle{Star Formation in Sh\,2-142}
\shortauthors{Sharma et al.}

\graphicspath{{./}{figures/}}
\usepackage{hyperref}

\begin{document}
\author[0000-0003-1634-3158]{Tanvi Sharma}
\affiliation{Institute of Astronomy, National Central University, 300 Zhongda Road, Zhongli 32001 Taoyuan, Taiwan}

\author[0000-0003-0262-272X]{Wen Ping Chen}
\affiliation{Institute of Astronomy, National Central University, 300 Zhongda Road, Zhongli 32001 Taoyuan, Taiwan}
\affiliation{Department of Physics, National Central University, 300 Zhongda Road, Zhongli 32001 Taoyuan, Taiwan}

\author[0000-0002-0151-2361]{Neelam Panwar}
\affiliation{Aryabhatta Research Institute of Observational Sciences (ARIES), Manora Peak, Nainital 263 002, India}

\author[0000-0002-3904-1622]{Yan Sun}
\affiliation{Purple Mountain Observatory, Chinese Academy of Sciences, 10 Yuanhua Road, Nanjing 210033, China}

\author[0000-0003-0007-2197]{Yu Gao}
\affiliation{Department of Astronomy, Xiamen University, Xiamen, Fujian 361005, China}
\affiliation{Purple Mountain Observatory, Chinese Academy of Sciences, 10 Yuanhua Road, Nanjing 210033, China}

\title{Diagnosing Triggered Star Formation in the Galactic \ion{H}{2} region Sh\,2-142}

\begin{abstract}
Stars are formed by gravitational collapse, spontaneously or, in some cases under the constructive influence of  nearby massive stars, out of molecular cloud cores.  Here we present an observational diagnosis of such triggered formation processes in the prominent \ion{H}{2} region Sh\,2-142, which is associated with the young star cluster NGC\,7380, and with some bright-rimmed clouds as the signpost of photoionization of molecular cloud surfaces.  Using near- (2MASS) and mid-infrared (WISE) colors, we identified candidate young stars at different evolutionary stages, including embedded infrared sources having spectral energy distributions indicative of active accretion. We have also used data from our optical observations to be used in SEDs, and from Gaia EDR3 to study the kinematics of young objects.  With this young stellar sample, together with the latest CO line emission data (spectral resolution $\sim 0.16$~km~s$^{-1}$, sensitivity $\sim 0.5$~K), a positional and ageing sequence relative to the neighboring cloud complex, and to the bright-rimmed clouds, is inferred. The propagating stellar birth may be responsible, at least partially, for the formation of the cluster a few million years ago, and for the ongoing activity now witnessed in the cloud complex. 
\end{abstract}


%sec1
\section{Introduction} \label{sec:Introduction}

Massive stars, via their fierce radiation and stellar winds, have profound effect on the environments. The influence, on the one hand, could be destructive so as to disperse nearby molecular gas, quenching any further star formation activity. On the other hand, the OB stars may play a constructive role under certain circumstances \citep{1998ASPC..148..150E,2007ApJ...657..884L}, with their winds giving ``just the right touch'' to sweep up the surrounding gas and dust to a dense shell that eventually fragments to form a group of young stars \citep[the collect-and-collapse process,][]{1977ApJ...214..725E}. For a neighboring molecular clump, a luminous star may ionize the confronting surface layer, which manifests itself as a ``bright-rimmed cloud'' or a cometary globule, thereby sending inwards an implosive ionization shock front that squeezes the material and initiates cloud collapse, leading to formation of the next-generation stars \citep[the radiation driven implosion,][]{1989ApJ...346..735B}. More than spontaneous cloud collapse, such triggering mechanisms \citep{1989ApJ...342L..87S, 1991ApJS...77...59S, 2004A&A...426..535M} extend the starbirth processes in time, producing stars at differing epochs in a cloud complex, and in length scales, propagating the formation sequence from one cloud to another.

The effect of OB stars on triggering formation of next-generation stars in nearby molecular clouds would leave the following spatial and temporal prognostic trace \citep{2007IAUS..237..278C}: namely, (1)~the remnant cloud is surrounding (in the collect-and collapse scenario) or pointing to (if by the radiation driven mechanism) the OB stars; (2)~the recently formed stellar groups are aligned between the remnant cloud and the OB stars in an age sequence, with the younger population being nearer to the bright cloud rim, whereas the youngest ones, i.e., protostars, if any, are being formed inside the cloud.
%{2007IAUS..237..278C} Chen, W.~P., Lee, H.~T., \& Sanchawala, K.\ 2007, Triggered Star Formation in a Turbulent ISM, 237, 278

The target of our study, the cloud complex Sh\,2-142, shown in Figure~\ref{fig:my_label}, presents a likely case of triggered star formation. The \ion{H}{2} region Sh\,2-142 was first recognized by Caroline Hershel and later cataloged by Stewart Sharpless \citep{1959ApJS....4..257S}. The principal ionizing star, DH\,Cep, is a spectroscopic binary (HD\,215835, O5.5V$+$O6V) associated with the cluster NGC\,7380 (C2245+578 in IAU nomenclature, $\alpha$= 22:47:16 , $\delta$= +58:07:30, J2000.0), which is at a heliocentric distance of 2.6~kpc and has an age of 4~Myr \citep{2011AJ....142...71C}. This young, open cluster also fosters a pre-main-sequence (PMS) population  \citep{1983A&AS...51..235B,2011AJ....142...71C,2016MNRAS.456.2505L}. The natal cloud embracing the cluster in the south and eastern side was revealed by molecular line observations \citep{1978A&A....70..769I,1989ApJS...70..731L}.  Furthermore,  \citet{1988ApJ...332.1030J} disclosed two kinematically and spatially separated cores within the complex. The gas in the region has been dispersed by stellar winds and radiation, yet remains interacting with grouped  young stellar objects (YSOs) \citep{2012ApJ...744..130K} and embedded far-infrared point sources.

The collation of far-infrared observations, H$\alpha$ imaging, radio continuum and molecular line data  \citep{1987ApJ...320..258S, 1984ApJ...283..640J,1985ApJ...298..596J,1994A&A...283..963C} has revealed several peaks of density enhancement of molecular and ionized gas, thus uncovering the presence of peripheral bright rims that are likely to harbor very young stars. Indeed a few bright-rims are present, including the extensively studied BRC\,43, known to associate with embedded sources, which serves as a potential site for triggered star formation by the "radiatively driven implosion" mechanism   \citep{1991ApJS...77...59S, 2008A&A...477..557M, 2010MNRAS.408..157M}.

In this paper we investigate the region to diagnose the star formation activity for the overall Sh2-142 cloud complex, by associating the young population with the star cluster, as well as with the molecular and ionized gas. Combining near- and mid-infrared data, we update the list of young stars in the whole complex, including those still accreting mass, and collectively analyze their  location and relative ages, to gain insight into the influence of massive stars on the star-formation history in neighboring molecular clouds.  In the following, Section~\ref{sec:Observations and Data} describes the data used in this work, archival or otherwise acquired by ourselves, leading to Section~\ref{sec:analysis} where the evolutionary status of the young objects is determined accordingly.  Section~\ref{results} then presents the kinematics of young stars, and subsequently the interplay between parental molecular gas and YSOs.   In section~\ref{discussion} we discuss supporting evidence for, and limitations and uncertainties in our arguments, of influential star formation in this region.

%sec2
\section{Observations and Data Sources} \label{sec:Observations and Data}

\begin{figure*}[htb!]
    \centering
    \includegraphics[width=0.8\textwidth]{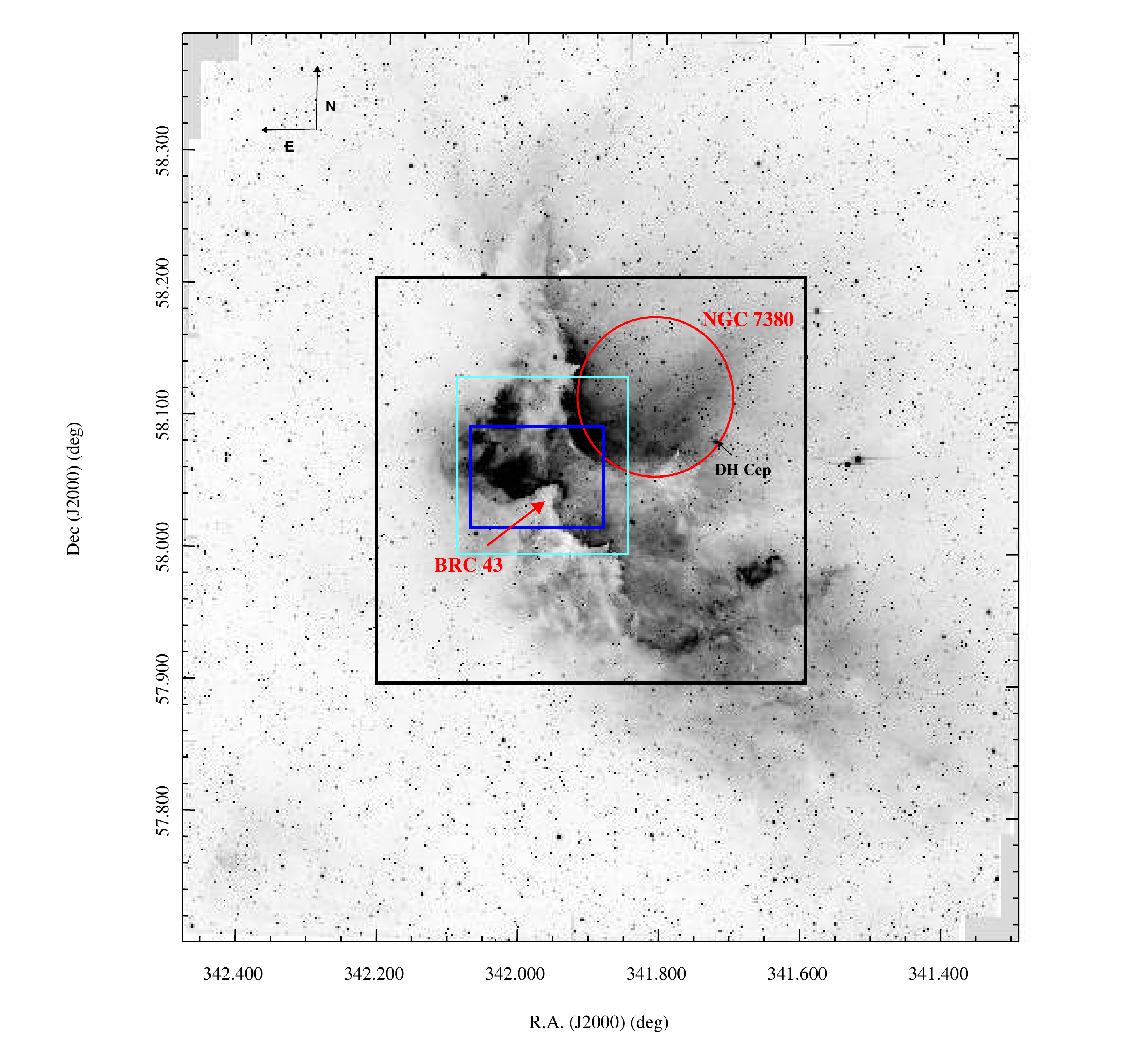}
    \caption{The Sh\,2-142 complex of our study (color-inverted H$\alpha$ image), where near- and mid-infrared data from 2MASS and WISE are analysed. The black square marks the region in which deep optical $BVI$ photometry from 1.3m DFOT is available, and cyan square  for 2m HCT, whereas the blue square is the region near BRC\,43 within which H$\alpha$ emission stars have been identified. The location of the cluster NGC\,7380 is shown by a red circle and the principal ionizing star (DH\,Cep) for the region is labeled. The centre for this  image is at $\alpha$=22:47:33.3, $\delta$=+58:02:56.8 (J2000).}
    \label{fig:my_label}
\end{figure*}

Data used in this study consist of archival photometry in the infrared wavelengths, optical broadband imaging, and molecular line observations.  For infrared data, we have explored an area of $\sim~38\arcmin\times 41\arcmin$ around NGC\,7380 covering the entire cloud complex, exhibited in  Figure~\ref{fig:my_label}, by exploiting the archival Two Micron All Sky Survey (2MASS) in $J$ (1.25~\micron), $H$ (1.65~\micron) and $Ks$ (2.15~\micron) bands \citep{2003yCat.2246....0C}, comprising a total of 18316 point sources. In addition, the ALLWISE catalog \citep{2013yCat.2328....0C} is accessed for photometry at longer wavelengths (at 3.4, 4.6, 12 and 22\micron, respectively, named as $W1$, $W2$, $W3$ and $W4$ bands), appropriate for embedded sources. This data set consists of 9457 sources.

CCD imaging Observations were carried out on 2018 October 6, using the Ritchey-Chretien Cassegrain 1.3-m telescope at Devasthal, Nainital, India, equipped with a  back-illuminated, thermoelectrically cooled CCD camera assembled by ANDOR, with $2048\times2048$ pixels. The $BVI$ images were processed following the standard procedure with bias and flat-field corrections. Point spread function photometry was then performed using the IRAF/DAOPHOT-II analysis package. The Landolt standard field SA\,92 was observed in the same night for photometric calibration. Additional deep  $BVI$ data was used toward BRC\,43 taken in 2006 October with the 2-m Himalayan Chandra Telescope (HCT) at Hanle, India, with the same data reduction procedure. The HCT observations are limited only to the BRC\,43 region, and for overlapping regions by both telescopes, we adopted the results by the 1.3~m for the bright targets ($V<17$)~mag and by the 2~m for fainter ($V\geq17$) ones.

The extinction coefficients in different filters for 1.3-m: $K_{U}=0.516\pm0.088$, $K_{B}=0.252\pm0.019$, $K_{V}=0.153\pm0.016$, and $K_{I}=0.083\pm0.017$. Transmission coefficient values are as follows:

$$U-B=0.938(U_{0}-B_{0})-1.390$$
$$B-V=1.215(B_{0}-V_{0})-0.847$$ 
$$V-I=0.917(V_{0}-I_{0})+0.211$$
$$V-V_{0}=-0.098(B-V)-2.282$$ 
$$B-B_{0}=0.031(U-B)-2.92.$$
The extinction coefficients, transmission coefficients, and zero points for the HCT are taken from \citet{2009MNRAS.396..964C}. 

We have used the Gaia EDR3 data \citep{2020yCat.1350....0G} to analyze the kinematics of the young population in the same region as the archival IR data and has 48847 sources towards Sh2-142, and utilize this data for parallax and proper motions.

%fig01

%\subsubsection{CO Isotope Emissions}

The molecular gas is traced by $^{12}$CO, $^{13}$CO and C$^{18}$O emissions, all of $J=1-0$ transitions, observed with the 13.7-m radio telescope at Delingha, China. The data are part of the long-term effort by the Purple Mountain Observatory, since nearly 10 years ago to map the Galactic plane \citep{2019ApJS..240....9S}. The data format is in three-dimensional cubes with two spatial axes (Galactic longitude and latitude) and one spectral axis (LSR velocity). For the $^{12}$CO emission at 115~GHz, the spectral resolution (channel width) is $\sim0.16$~km~s$^{-1}$ with a typical sensitivity of about 0.5~K. For $^{13}$CO and C$^{18}$O near 110~GHz, the resolution and typical sensitivity are, respectively, 0.17~km~s$^{-1}$ and $\sim0.3$~K.  CO data toward Sh\,2-142 have either sub-arcmin spatial resolution, or high resolution (30\arcsec) but limited only to the vicinity of BRC\,43 \citep{1988ApJ...332.1030J}.   Our molecular line data cover the entire complex at an angular resolution of $\sim$50\arcsec\  with a uniform sensitivity.    

%sec3
\section{Identification of Young Stellar Objects} \label{sec:analysis}

The young stellar sample gathered for this study include those known to display emission lines in the spectra, and those that have characteristic infrared radiation in excess of photospheric radiation, originated from circumstellar dust emission, diagnosed by near-infrared and mid-infrared colors.  

\subsection{Emission-line and Near-Infrared Excess Stars}

PMS stars are known to exhibit the H$\alpha$ line in emission in their spectra as a sign of chromospheric activity or circumstellar accretion, both indicative of stellar youth. We adopted the sample of 14 H$\alpha$ emission stars compiled by \citet{2002AJ....123.2597O} that are mostly located outside or near the tip of BRC\,43, as illustrated by blue square in Figure~\ref{fig:my_label} of the spatial coverage of our study. The equivalent width listed in their work allows distinction between classical T~Tauri stars (CTTS, with  $EW >10$~\AA) and weak-lined T~Tauri stars (with $EW <10$~\AA), the latter being relatively more evolved so as to have largely cleared their inner disks.  This sample is limited in spatial coverage and, being detected in optical wavelengths, is also biased against embedded sources.  Relatively evolved stars, such as magnetically active dwarfs (dMe) are potential contaminants.

% \subsubsection{Near-Infrared Excess Stars}

\begin{figure*}[htb!]
\centering
\includegraphics[width=0.8\textwidth]{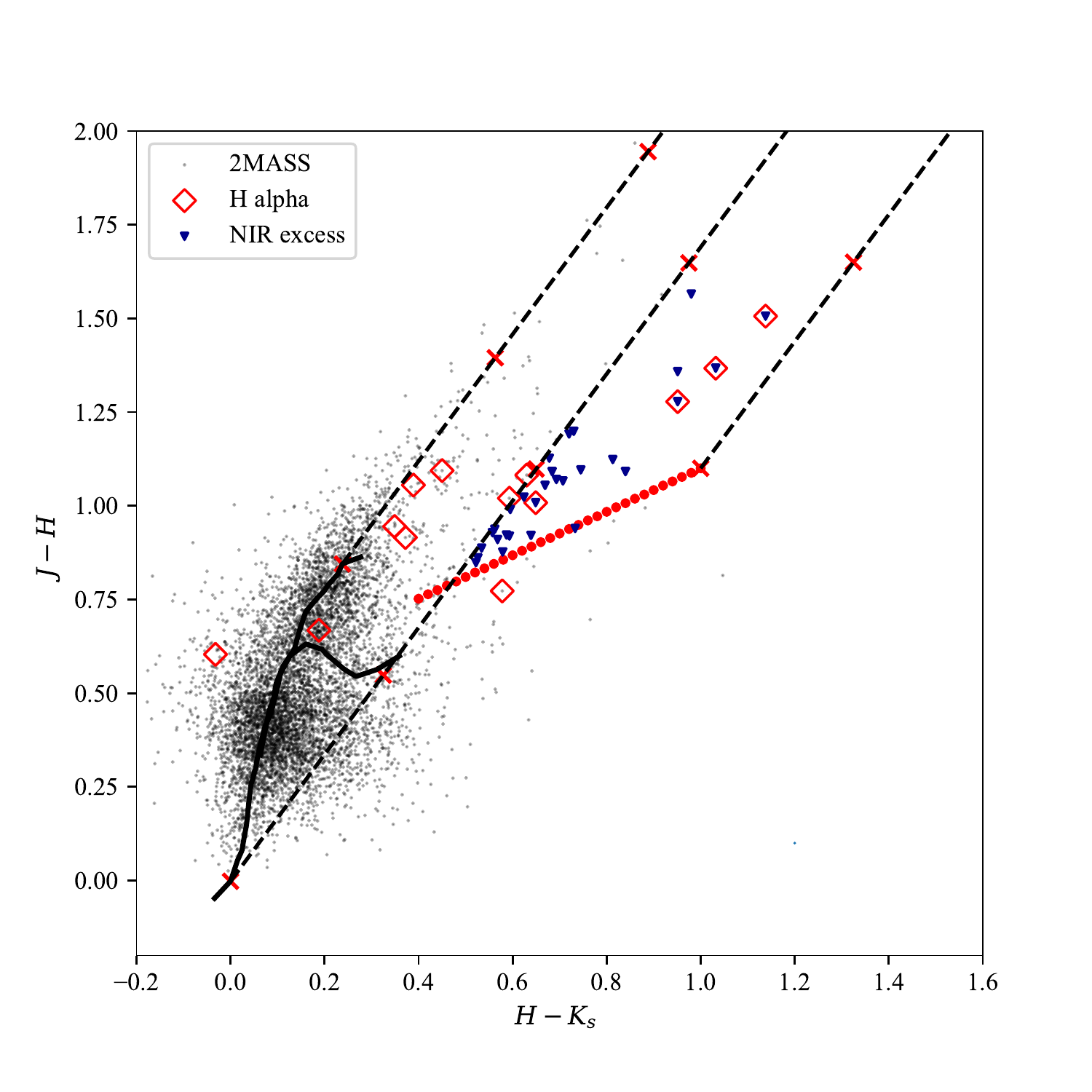}

\caption{Near-infrared 2MASS $J-H$ versus $H-K_s$ color-color diagram. The gray dots represent 2MASS measurements, whereas the black curves represent unreddened main sequence and giant star loci \citep{1988PASP..100.1134B}. The intrinsic locus of classical T~Tauri stars is marked by the dotted red line \citep{1997AJ....114..288M}. All the data are transformed to the CIT system \citep{2001AJ....121.2851C}. Reddening vectors, shown as dotted black lines, adopting an interstellar law \citep{1981ApJ...249..481C} of $A_J/A_V=0.265$, $A_H/A_V=0.155$ and  $A_K/A_V=0.090$, are drawn from the extreme tips of each locus, with the red crosses on each vector marking an increment of $A_V=5$.  The red open diamond symbols mark H$\alpha$ stars, whereas the blue filled symbols show infrared excess objects diagnosed in this diagram.}
\label{fig:jhk_ccd}
\end{figure*}

Excess of infrared emission is the consequence of absorption of starlight in UV and visible by circumstellar grains, which then reradiates at longer wavelengths. The position in a color-color diagram essentially represents different slopes in the spectral-energy distributions (SEDs), with $\alpha = d\log (\lambda$ $F_{\lambda})/d\log (\lambda)$, where $\lambda$ is the wavelength, $F_\lambda$ is the flux density, and the slope  $\alpha$ serves as the proxy of the amount of thermalized dust, and hence the evolutionary stages of YSOs. Here we use a $J-H$ vs $H-K_S$ color-color diagram capable of qualifying the level of infrared excess and distinguishing this from interstellar reddening.  Of the 18316 2MASS sources in our field, 6901 have the photometric quality flag ``AAA'', i.e., with an uncertainty less than $0.1$~mag in all of the three bands. These are the stars used in our analysis for the 2MASS sample. Comparison of the observed 2MASS colors with the intrinsic loci of main sequence stars and of giants, including added effects of interstellar reddening, leads to classification, in progressively increasing infrared excess (proxy of amounts of circumstellar dust), of classical T~Tauri (solar-type) candidates, Herbig Ae/Be (intermediate-mass PMS) candidates, or protostars.  In Figure~\ref{fig:jhk_ccd}, the rightmost trapezoid enclosing reddened T~Tauri stars, but excluding main sequence stars experiencing normal interstellar reddening, signifies the region within which young stars are likely located.  A total of 29 such near-infrared excess objects, namely classical T~Tauri candidates, were identified, these are shown in Figure~\ref{fig:jhk_ccd}, together with H$\alpha$ emission sources.  

%fig02

\subsection{Mid-infrared Excess Stars}

\begin{figure*}[htb!]
    \centering
    \includegraphics[width=0.8\textwidth]{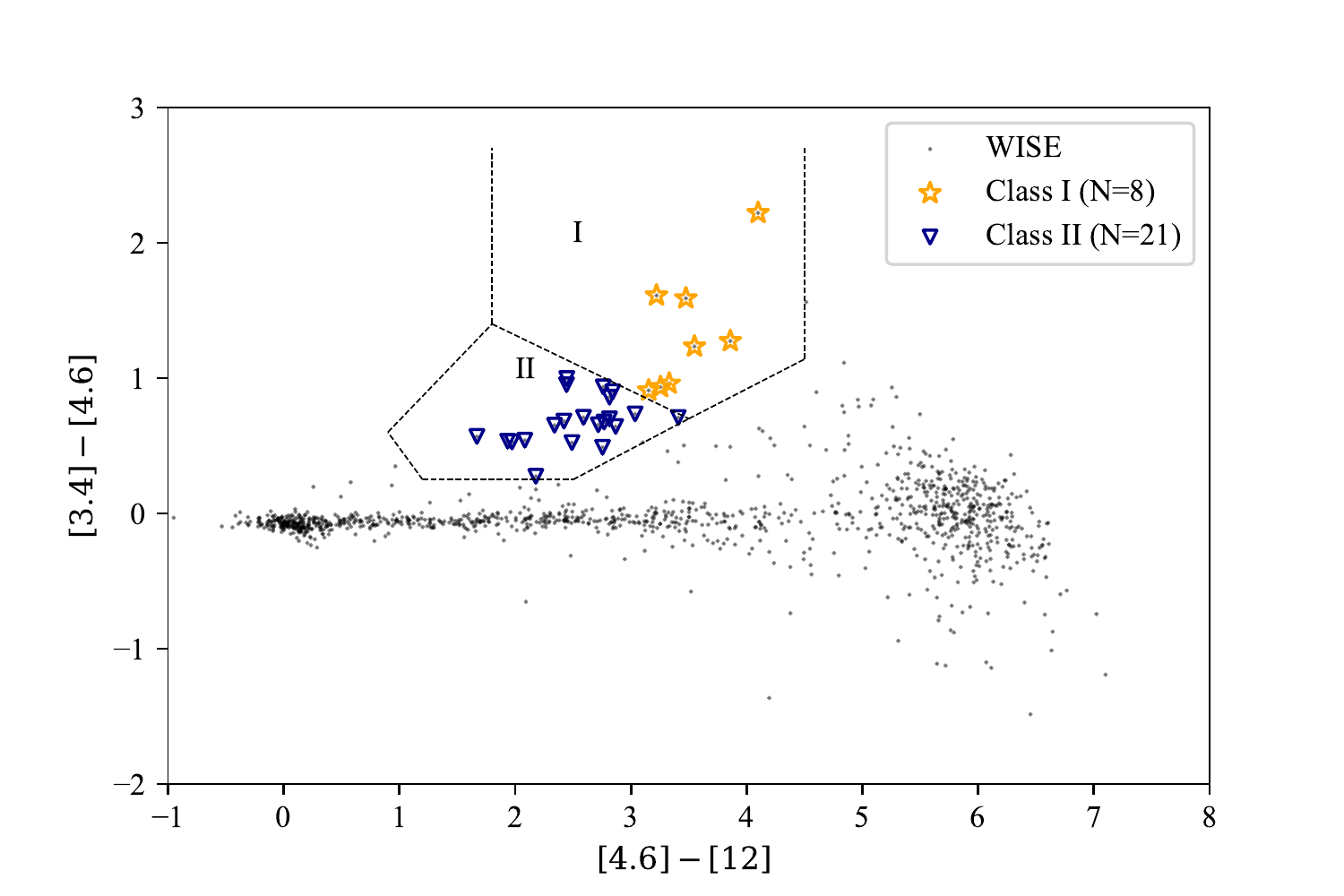}
    \caption{The WISE mid-IR color-color diagram, from 3.4~\micron\ to 12~\micron, to identify Class~I objects (represented by orange asterisks) and Class~II objects (open triangles in blue), following the selection criteria in \citet{2014ApJ...791..131K}.  Grey dots mark all 1846 sources with photometric quality flag ``AAA''.}
    \label{fig:wise_ysos}
\end{figure*}

Deeply embedded stars may escape 2MASS detection; these are then selected from archival WISE data.  Of the 9457 sources in the region, 1846 have photometric uncertainties less than $0.1~$mag in 3.4~$\micron$ ({\it W1}), 4.6~$\micron$ ({\it W2}), and 12~$\micron$ ({\it W3}), with the quality flag 'AAA'.  We adopted the scheme of \citet{2014ApJ...791..131K}, i.e., with color cuts from {\it W1} to {\it W3} for different YSO classes, while winnowing out extragalactic contaminants such as active galactic nuclei or star-forming galaxies. Using ({\it W1})$-$({\it W2}) versus ({\it W2})$-$({\it W3}), we retrieved eight Class~I and 21 Class~II objects, as illustrated in Figure~\ref{fig:wise_ysos}. Three out of eight Class~I objects have reliable detection in every 2MASS band, each with consistency of near-IR excess. The remaining five do not have reliable 2MASS measurements, plausibly indicative of elevated extinction. Out of the 21 WISE-selected T~Tauri candidates, 13 appear to also exhibit an excess of near-IR ($J-H \ga 0.75$ and $H-Ks \ga 0.4$ in Figure~\ref{fig:jhk_ccd}). Four of the 21 WISE-selected T~Tauri candidates are also common to NIR selected CTTS sample.

\begin{figure*}[htb!]
    \centering
    \includegraphics[width=0.8\textwidth]{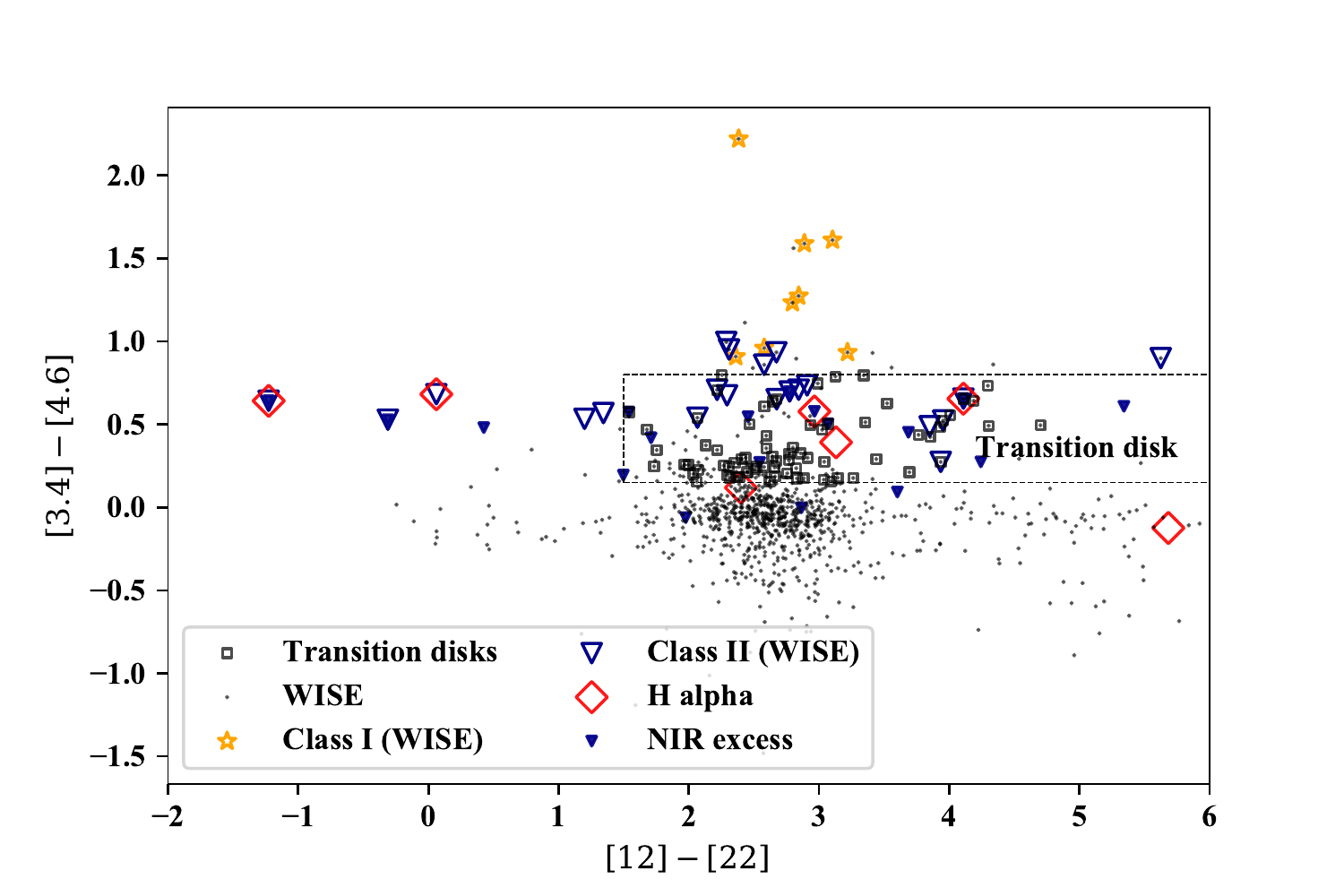}
    \caption{The WISE color-color diagram, similar to Figure~\ref{fig:wise_ysos} but here extends to 22~\micron, where transition disk are likely to be found is marked in this figure.  The Class~I and Class~II objects identified in Figure~\ref{fig:wise_ysos} are marked with the same symbols. Note that Class~II objects do not stand out distinctly in this diagram.  Black dots represent data points with a photometric error $<0.1$~mag in each of the four WISE bands. }
    \label{fig:tr_disk}
\end{figure*}

More embedded sources need to be investigated at longer wavelengths.  Figure~\ref{fig:tr_disk} presents the $W1-W2$ versus $W3-W4$ color-color diagram, in which transition-disk objects stand out, exhibiting excess in 10--20~\micron.  By contrast, Class~II objects with prominent near-infrared excess can no longer be selected distinctly  in this diagram as in Figure~\ref{fig:jhk_ccd} or Figure~\ref{fig:wise_ysos}. 
Note that in Figure ~\ref{fig:tr_disk}, a few H$\alpha$ stars, candidate CTTSs are also common to transition-disk objects, which are a subset of T-Tauri stars with the inner disks being cleared; that is, they are in an evolutionary phase between Class~II and Class~III objects, and the common ones are included in our young stellar sample.  The analysis above demonstrates the necessity of using a combination of colors, data quality permitting, with a wider range of wavelength coverage to acquire a more complete YSO sample.  
The catalog of PMS candidates is presented in Appendix (Table \ref{Tab:all_list}).

%fig03

%fig04

\begin{figure}[htbp!]
\centering

  \includegraphics[width=\columnwidth]{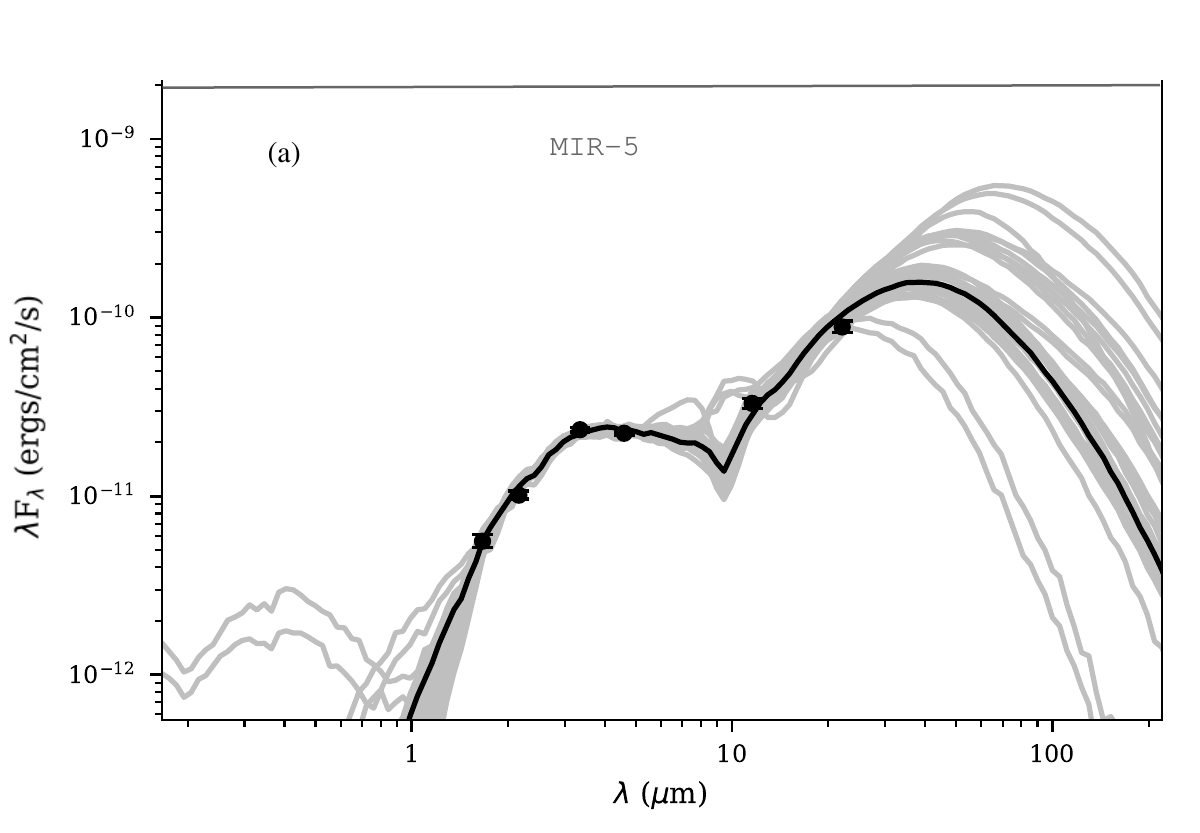}
  \includegraphics[width=\columnwidth]{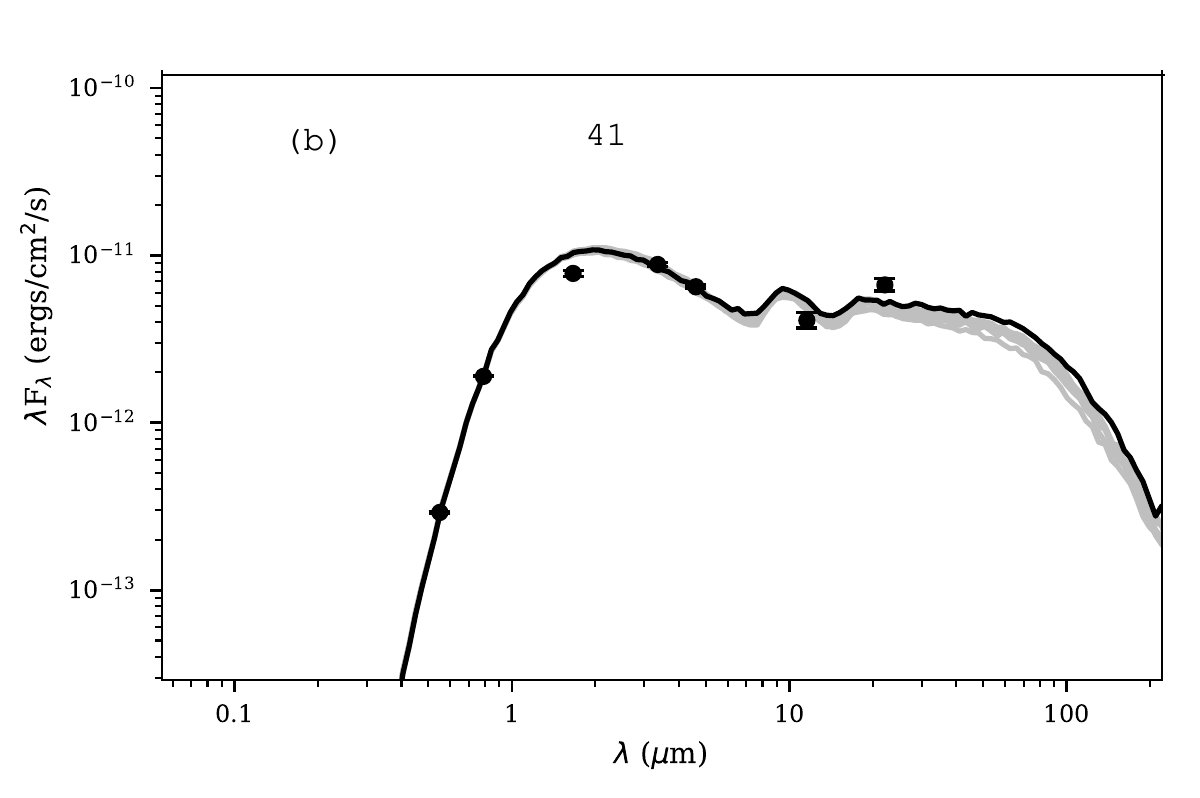}
\caption{
  Example SEDs of (Top)~a Class~I and (Bottom)~a Class II object. In each case, the source identification number is labeled.  The black curve represents the best-fit model in the least-squares sense, and the grey curves indicate models with the range of reasonably good fits, i.e., with $\chi_{min}^{2} -\chi^{2} <3N$, where $N$ is the number of observational data points. The best-fit parameters of the Class~I object, MIR-5, are listed in Table~\ref{Tab:Table_sed}.
         }
    \label{fig:seds}
\end{figure}

We matched counterparts within a 3$\arcsec$ radius with the $BVI$ (this work), 2MASS, and WISE sources, to construct the SED of each source for further investigation.   The SED fitting was done with the tool developed by \citet{2007ApJS..169..328R}, which uses a pre-computed grid of 20,000 young stellar models to constrain the parameters for the central object (e.g., mass, temperature, age) and for the envelope (e.g., disk mass and accretion rate). The number of observational data points, and respective wavelength coverage, dictates the reliability of the constraints on the parameters. We required a minimum of four data points in an SED, and allowed in the fitting the extinction ($A_V$) to vary from 1.5~mag (minimum foreground extinction toward the cluster) to 15~mag, and the distance to vary from 2.2~kpc to 3.5~kpc, encompassing the literature values \citep{1976A&A....49...57G, 1971A&A....13...30M,  1994A&A...283..963C}. Figure~\ref{fig:seds} illustrates distinct SED behavior of an  example of a Class~I, with an ascending SED with wavelength, versus that of a Class~II object.  The best-fit parameters such as the disk mass ($M_{\rm Disk}$), central object mass ($M_{*}$), temperature ($T_*$), adopting the approach given in \citet{2012ApJ...755...20S} using weighted mean and the standard deviation, are presented in Table \ref{Tab:Table_sed} for Class~I objects.  The results suggest that most Class~I objects have ages $\la1$~Myr, consistent with them being young and embedded.

\begin{deluxetable*}{cLLRRRRRRR}[!]

\tabletypesize{\footnotesize}

\tablecaption{ SED fitting parameters for Class~I YSOs}

\tablenum{1}

\tablehead{\colhead{ID} & \colhead{$\alpha$~ (J2000)} & \colhead{$\delta$~ (J2000)} & \colhead{$[3.4]-[4.6]$} & \colhead{$[4.6]-[12]$} & \colhead{Age} & \colhead{$M_{*}$} & \colhead{$T_{*}$} & \colhead{$M_{\rm Disk}$} & \colhead{$\dot{M}$} \\
\colhead{} & \colhead{  (deg)} & \colhead{  (deg)} & \colhead{ (mag) } & \colhead{ (mag) } & \colhead{ (Myr) } & \colhead{($M_\sun$)} & \colhead{ ($10^3$~K)} & \colhead{ ($M_\sun$) } 
& \colhead{($10^{-5} M_\sun$~yr$^{-1}$})}
\startdata
MIR-1 & 341.440025 & 57.847372 & 1.590 & 3.475 & 0.002 & 0.623 & 3.776 & 0.006 & 17.637 \\
MIR-2 & 341.575326 & 57.989582 & 1.611 & 3.221 & 0.937 & 0.850 & 4.842 & 0.006 & 10.649 \\
MIR-3 & 341.642634 & 57.995816 & 1.234 & 3.549 & 0.021 & 2.455 & 4.09 & 0.033 & 466.956 \\
MIR-4 & 341.943429 & 58.130137 & 1.274 & 3.859 & 0.015 & 4.73 & 4.224 & 0.036 & 67.637 \\
MIR-5 & 341.969577 & 58.131370 & 0.934 & 3.256 & 0.713 & 3.26 & 5.864 & 0.063 & 16.305 \\
MIR-6 & 341.981224 & 58.160661 & 2.221 & 4.099 & 2.470 & 3.661 & 12.547 & 0.025 & 92.607 \\
MIR-7 & 341.898887 & 58.153404 & 0.960 & 3.331 & 0.500 & 4.891 & 7.9 & 0.000 & 50.080  \\
MIR-8 & 341.93185  & 58.328075 & 0.909 & 3.153 & 0.258 & 0.343 & 2.961 & 0.004 & 0.786 \\
\enddata
\end{deluxetable*}
\label{Tab:Table_sed}

%fig05

%sec4

\section{Properties of Young Stars }  \label{results}

%sec4.1
\subsection{Kinematics of the Young Stellar Sample}

\begin{figure*}[htb!]
    \centering
    \includegraphics[width=\textwidth]{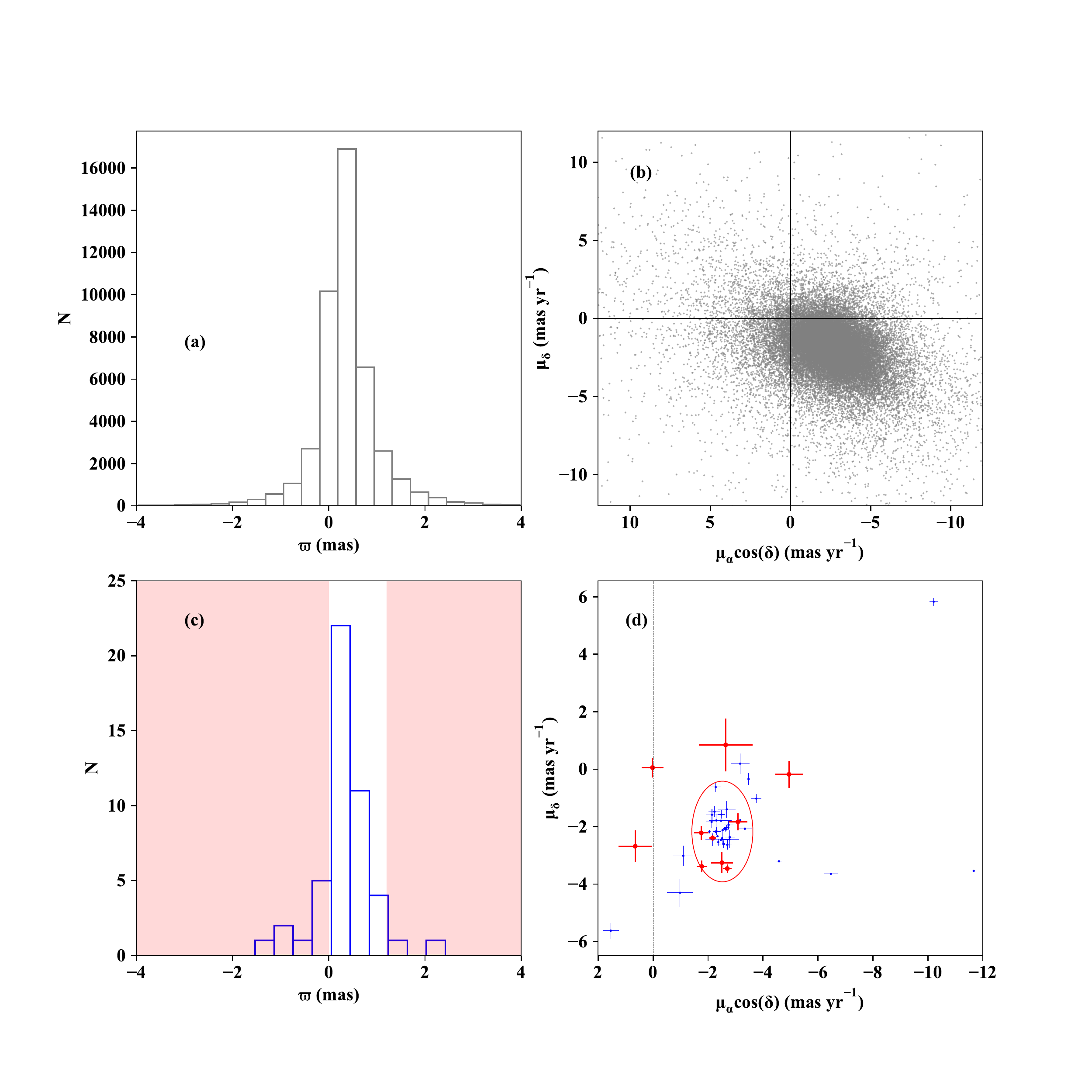}
    \caption{(a) Gaia EDR3 parallax and (b) proper motion vectors for all sources, whereas (c) and (d) show the same for the young stellar sample (in blue).  In (d) the ellipse depicts 4 times median-absolute deviation of the young sample, centered on the mean proper motion for the cluster NGC\,7380.  Those stars in the shaded region in (c) having $\varpi \leq0$ or $\varpi \geq 1.2$~mas, signifying uncertain parallax membership, are marked by red symbols in (d). 
            }
 \label{fig:kinematics_GAIA}
\end{figure*}

We cross-matched the list of young objects in the region with Gaia EDR3 sources \citep{2020yCat.1350....0G} within a coincidence radius of 2\arcsec, resulting in two out of eight candidate protostars (Class~I), and 46 out of 54 PMS objects having parallax and proper motion measurements. To our knowledge, no radial velocity information is available in the literature for any of the matched sources. The Gaia EDR3 ID and measurements are listed in the YSO catalog presented in Appendix. The distributions of parallax and proper motion values are shown in Figure~\ref{fig:kinematics_GAIA}.  
 The region of our study is nebulous, rendering accurate astrometric estimation, with which parallax and proper motion are derived, difficult. This is evident by a noticeable fraction of negative parallax values, including some of our young candidates.   The median motion of young stellar sample is  ($\mu_{\alpha} \cos\delta, \mu_{\delta}) = (-2.52, -2.16)$~mas~yr$^{-1}$ which is also close to the average proper motions of the cluster NGC\,7380, ($\mu_{\alpha} \cos\delta, \mu_{\delta}) = (-2.517\pm0.131 , -2.144\pm0.131)$~mas~yr$^{-1}$ \citep{2018A&A...618A..93C}. Those objects that deviate more than 4 times the median-absolute deviation, therefore being inconsistent with the cluster's motion
are noted in Figure~\ref{fig:kinematics_GAIA}(d) and in Table~\ref{Tab:all_list} in Appendix.   In Figure~\ref{fig:kinematics_GAIA}(c) those young stars having $\varpi \leq0$ or $\varpi \geq 1.2$~mas, which would have been excluded of membership however exhibit a grouping in proper motion (Figure~\ref{fig:kinematics_GAIA}(d))
consistent with that of the cluster NGC\,7380.

There are objects with large deviations from the cluster's average proper motion, yet with youth indicators such as H$\alpha$ or infrared excess. To name a few, e.g., Star~34, a near-IR candidate exhibiting outlying proper motion of $(-10.2, 5.8)$~mas~yr$^{-1}$, a parallax of $1.06\pm0.14$~mas and a distance of $0.94_{-0.11}^{+0.13}$~kpc is likely not a member. Other examples include the H$\alpha$ star~9, and the candidate young star~37, both having motions inconsistent with membership. Four objects, namely stars~6, 23, 26, and 43, have negative parallax values and their proper motions are also not in accord with membership. 
%For instance, Star~6 is an H$\alpha$ emission-line star with negative parallax $-1.53\pm0.55$~mas, making it not possible to compute heliocentric distance by direct inverting parallax.  Its geometric distance $4.81_{-1.86}^{+1.46}$~kpc \citep{2021AJ....161..147B} manifests a large error. 

\begin{figure*}[htb!]
    \centering
    \includegraphics[width=\textwidth]{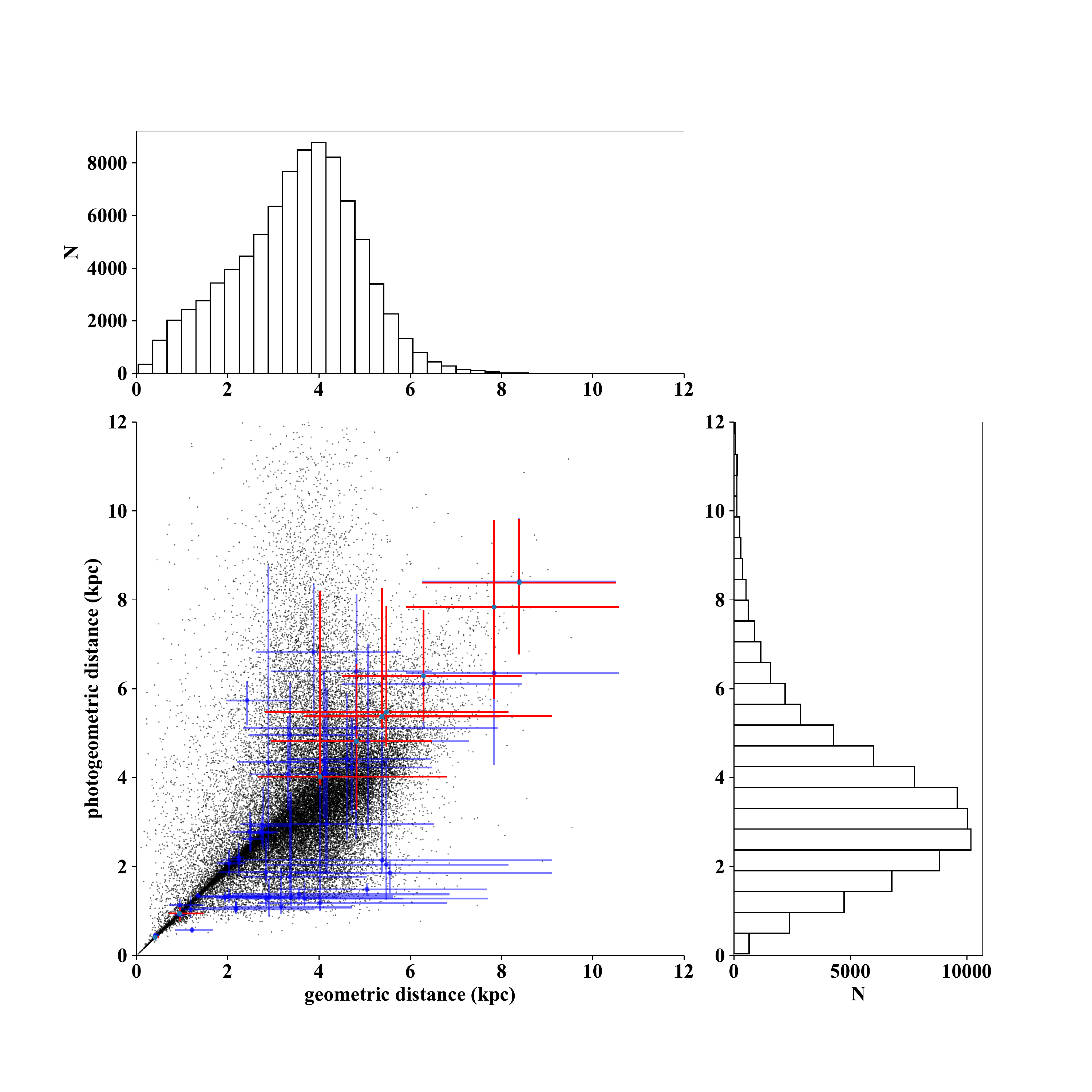}
    \caption{Geometric vs photogeometric distance estimated by \citet{2021AJ....161..147B}, and the respective number distribution.  The black points represent all sources, whereas the blue symbols mark the YSO sample along with asymmetric error bars.  The red symbols show the YSOs as parallax outliers in the shaded region in Figure~\ref{fig:kinematics_GAIA}(c), also the red crosses in Figure~\ref{fig:kinematics_GAIA}(d). 
            }
 \label{fig:geo_vs_photgeo}
\end{figure*}

% To remedy the effects of large uncertianties in parallax for such  nebulous regions
To remedy the effects of negative parallax, \citet{2021AJ....161..147B} derived the Bayesian geometric distances by direction-dependent prior from a parallax value and its uncertainties.  Additionally these authors also estimate a photogeometric distance based on the color, apparent magnitude, and extinction of a source. We compare the distributions of both distance determinations, shown as Figure~\ref{fig:geo_vs_photgeo} for all Gaia sources and for the YSO sample in our field of study. It is evident that the errors in distance computation for the YSOs are substantially large, essentially for those with negative parallax values.

 %fig06

 %fig07

Notably, Star~11 \footnote[1]{ Gaia EDR3 ID 2007420160273312512} is plausibly one of the stars whose parallax data are affected by the copious \ion{H}{2} nebulosity. As remarked in  Table~\ref{Tab:all_list}, it is a known emission-line object \citep{2002AJ....123.2597O} and exhibits prominent IR excess. Its parallax $2.42\pm0.26$~mas, however, would place it at a heliocentric distance of $\sim413$~pc and a geometric distance of $409.1_{-40.7}^{+73.7}$~pc, making it a puzzling, isolated, foreground young star. Nevertheless, its proper motion remains consistent with being a member of the cluster. The above analysis thus indicates the primacy of proper motions over parallaxes in establishing the association of young objects to Sh2-142.

%sec4.2
\subsection{Interconnection between Young Stars and Gas }

\begin{figure*}[htb!]
    \centering
    \includegraphics[width=\textwidth]{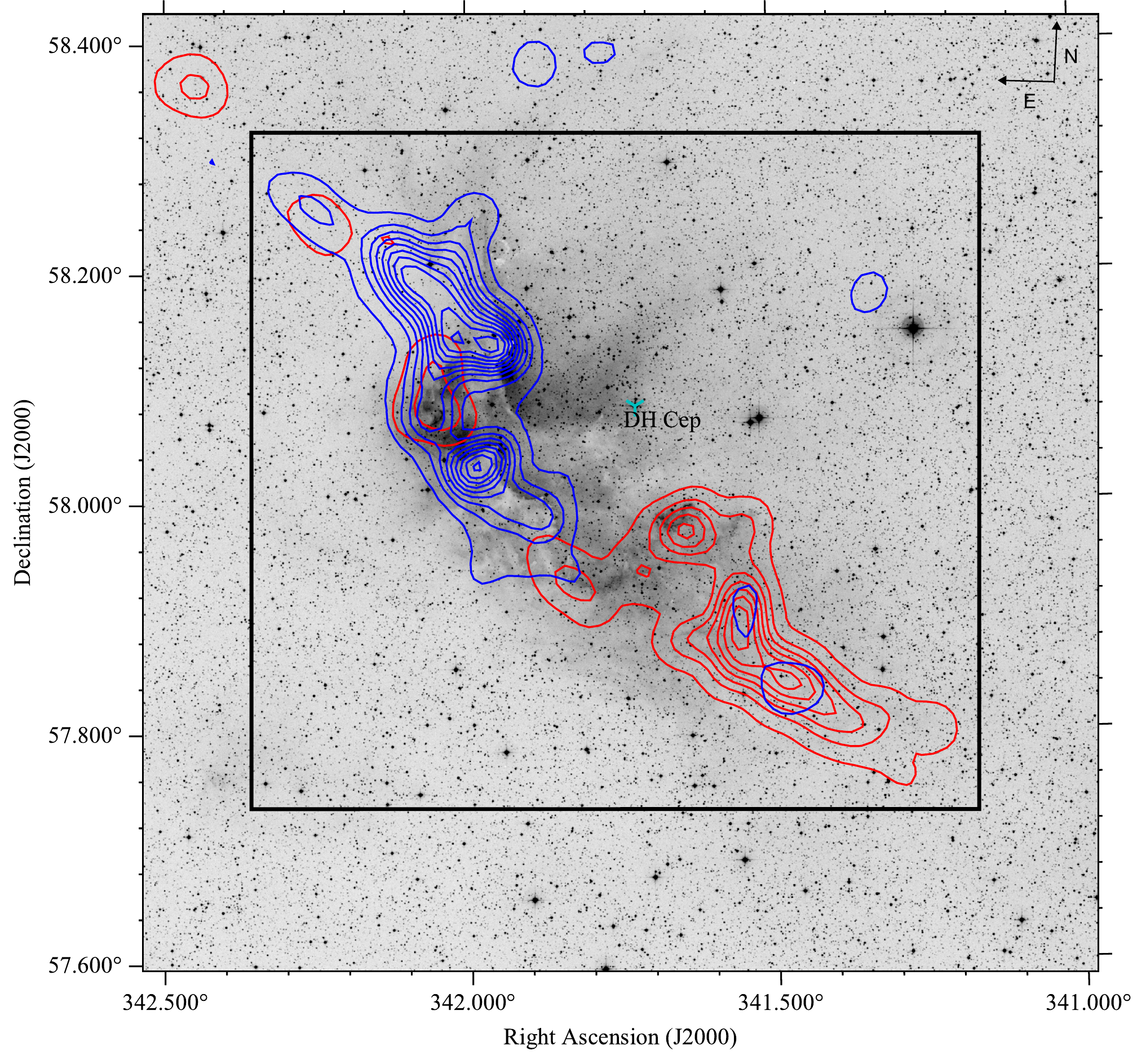}
    \caption{ 
     Integrated $^{12}$CO emission contours overlaid on the DSS2 red image, of the velocity range from $-45$~km~$s^{-1}$ to $-39$~km~s$^{-1}$ (in blue), and from $-39$~km~s$^{-1}$ to $-35$~km~s$^{-1}$ (in red). Contours are plotted from the minimum of 3$\sigma$, in steps of 3.5~K~km~s$^{-1}$. The southwest lobe is red shifted relative to the northeast lobe. The black square marks the boundary for which integrated CO spectrum is shown in the subsequent Figure~\ref{fig:spectrum}}
 \label{fig:integ_CO_ctr}
\end{figure*}

\begin{figure*}[htb!]
 \centering
\includegraphics[width=1.0\textwidth, angle=0]{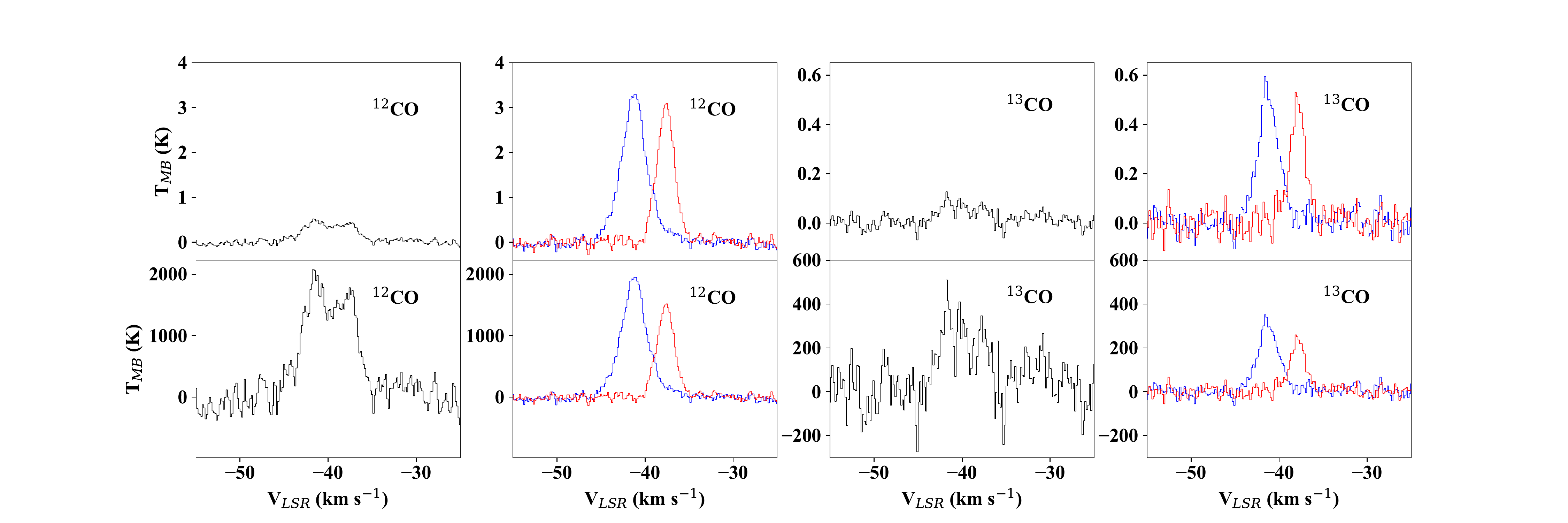}
\caption{
  The average (upper panels) and summed (lower panels) spectra between $-55$~km~s$^{-1}$ and $-25$~km~s$^{-1}$ for the $^{12}$CO and $^{13}$CO lines.  Each black curve shows the corresponding spectra within the black square area in Figure~\ref{fig:integ_CO_ctr}. The red and blue curves trace the outermost 3$\sigma$ contour for, respectively, the blue- and the red-shifted lobes outlined in Figure~\ref{fig:integ_CO_ctr}. 
  }
  \label{fig:spectrum}
\end{figure*}

Our CO data presented in Figure~\ref{fig:integ_CO_ctr} and \ref{fig:spectrum} evince spatial and kinematic structures of the molecular gas in the region. Spatially averaged and summed spectra, for $^{12}$CO and isotopic $^{13}$CO lines, clearly indicate two main lobes in the velocity range of $-55$ to $-25$~km~s$^{-1}$, bifurcating around $-39$~km~s$^{-1}$.   The $^{13}$CO lines are weaker, but consistently validate the kinematic and spatial segregation of the gas.  The C$^{18}$O emission is much weaker, hence not considered in our analysis.

The radial velocity of DH\,Cep, $-35.4\pm2.0$~km~s$^{-1}$ \citep{1953GCRV..C......0W}, matches to that of the average value of NGC\,7380, $-34.13\pm 6.09$~km~s$^{-1}$ \citep{2005A&A...438.1163K, 2017A&A...600A.106C}.  Jointly these then agree with the line-of-sight velocity of the bulk molecular gas, thus establishing the physical connection of the gas, star cluster, and the main exciting star in the complex.

 %fig08

%fig09

%\subsubsection{Morphology and kinematics} \label{result}

Morphologically, the CO emission clearly embraces the cluster \citep{2011AJ....142...71C, 1989ApJS...70..731L}, with a northeast-southwest extension, mainly in two lobes comparable in spatial extents but distinguished in kinematics. The two velocity peaks are $\sim 4$~km~s$^{-1}$ apart, with each peak corresponding to a lobe, in agreement with what has been reported by \citet{1988ApJ...332.1030J}.  Similarity in peak velocities of the two lobes with well distinct space volumes suggests that the clouds are not independent. Slicing through the channel maps with a step of 1~km~s$^{-1}$ reveals no systematic trend indicative of an overall cloud rotation.  On the other hand, the lobes are projected to have a separation of $\sim25$~pc at 2.6 kpc, a length scale too long for a protostellar outflow. Typical length scales for molecular outflows are of order of a parsec \citep{1988ApJS...67..283L}. The two lobes therefore cannot be due to either cloud rotation or stellar outflows. 

The relative symmetry of the lobes, plus the small velocity gradient across the cloud, indicates that the  lobes are near the plane of the sky.  We propose that the two lobes are parts of the ``working surface'' of the stellar winds and ionizing shock fronts upon the compressed clouds. There are a couple of known O-type stars in the region \citep{2011AJ....142...71C}, notably the DH\,Cep binary system, which emits copious X-rays possibly due to powerful colliding winds \citep{2002A&A...388L..20P,2010NewA...15..755B}.  Such a phenomenon is seen also, for example, in Orion and Monoceros where massive stars, molecular gas, and photoionized gas interplay.

\subsection{Reddening and Ages}  

\begin{figure}[htb!]
    \centering
    \includegraphics[width=\columnwidth]{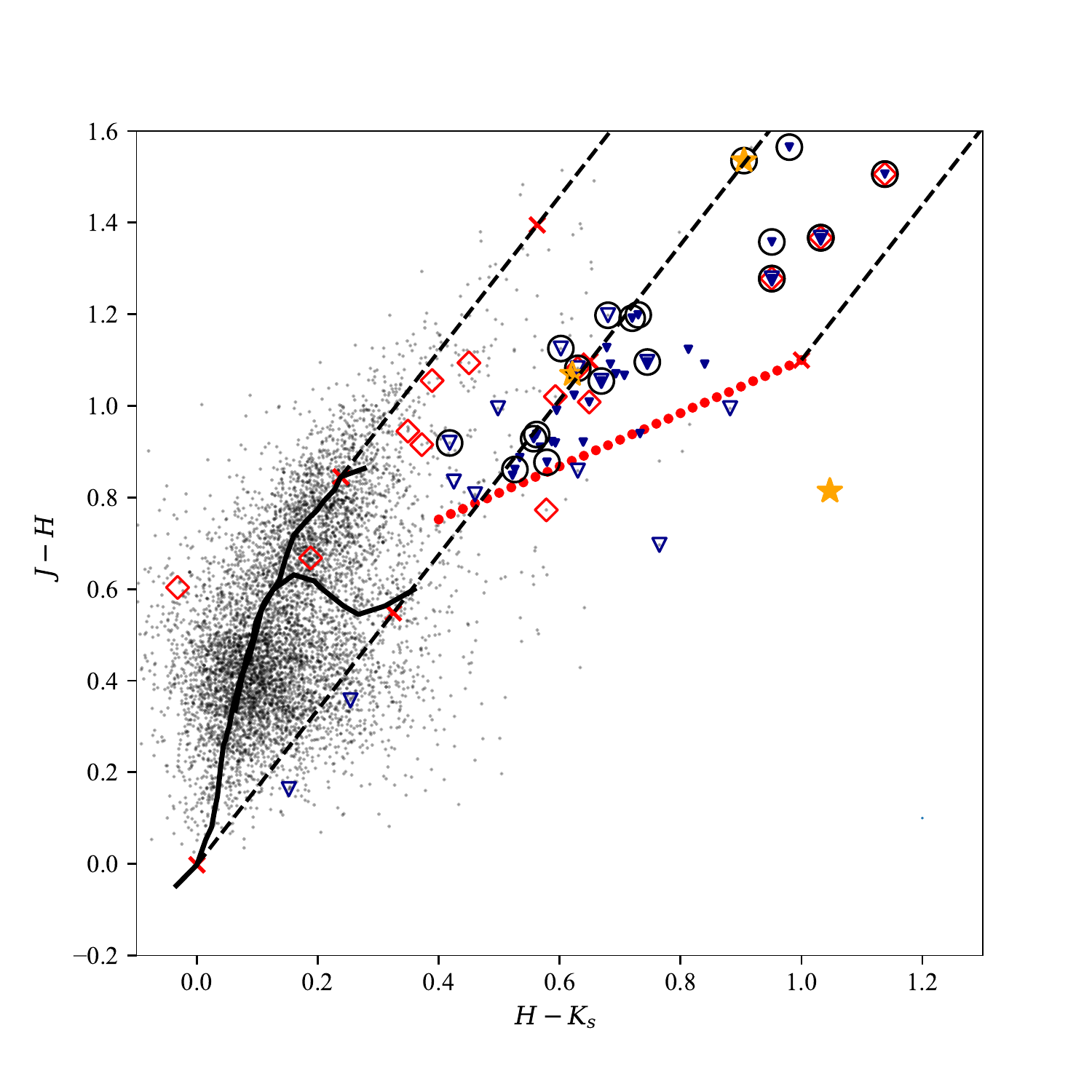}
    \caption{ $JHK$ color-color diagram for the H$\alpha$ stars (in red diamonds), and NIR stars (in filled blue triangles).  Also shown are 2MASS counterparts of some of the Class~I (orange asterisks), and Class~II (open blue triangles) sources identified with the WISE data (see Fig.~\ref{fig:wise_ysos}).
    }
    \label{fig:allccd}
\end{figure}

Of the H$\alpha$ stars (14), near- (29) and mid-infrared excess (29) objects, 38/62 are consistent with being classical T~Tauri stars in the NIR color-color diagram (see Figure~\ref{fig:allccd}). A few objects  displaying less IR excess yet with H$\alpha$ emission are likely evolved PMS, namely weak-lined T-Tauri stars. To estimate the extinction of each star, we tracked back in the diagram along the reddening vector from the observed colors to the intrinsic classical T~Tauri locus (the red line) or, for those outside the T-Tauri regime, to the main sequence (the black curve). This analysis obviously does not apply to heavily embedded objects as they are not detected by 2MASS.  

%fig10

\begin{figure}[htb!]
    \centering
\includegraphics[width=\columnwidth]{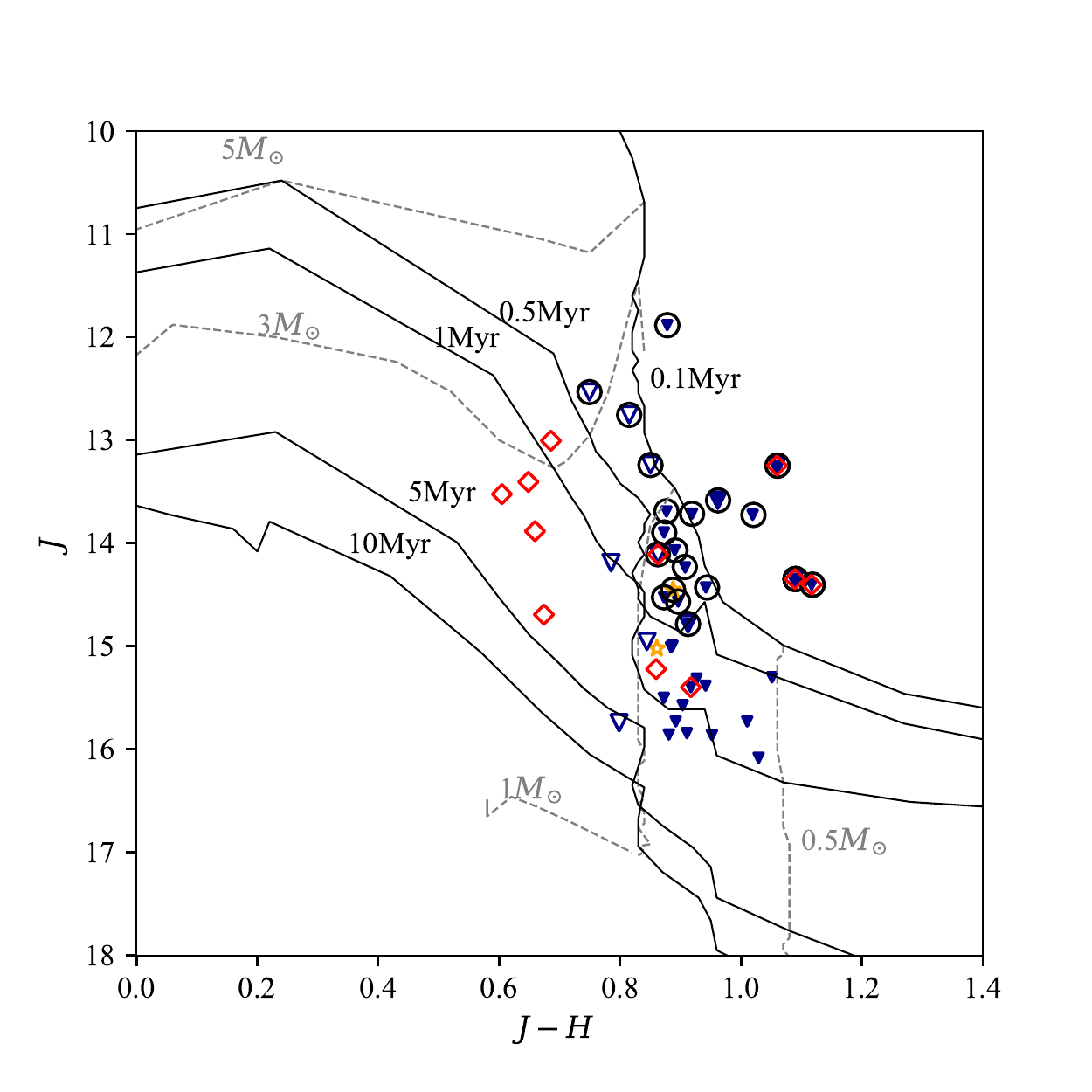}
    \caption{Dereddened 2MASS $J$ versus $J-H$ color-magnitude diagram for the young stars, with the same symbols as in Figure~\ref{fig:allccd}, 
    in comparison with PMS theoretical isochrones, and evolutionary tracks of \citet{2000A&A...358..593S}, corrected for a distance of 2.6~kpc, with the ages (solid curves) and masses (dashed curves) marked and labeled.  The observed $J$, and $J-H$ values are extinction corrected and dereddened according to the reddening law from \citet{1981ApJ...249..481C}.  The youngest objects are encircled in black. }
    \label{fig:final_cmd}
\end{figure}

Once the extinction is estimated, a young object is dereddened and placed in a color-magnitude diagram to infer its age and mass by comparison with theoretical isochrones, e.g., in the $J$ versus $J-H$ diagram as exhibited in Figure~\ref{fig:final_cmd}.  While isochrone ages may not be reliable in an absolute sense, it is seen that a few H$\alpha$ stars are relatively older than the near-IR excess sources. Most of the young population have masses $\la$ 1M$\sun$. The objects comprising the youngest group, with isochrone ages $\la 0.5$~Myr, are marked separately for further scrutiny.

%fig11

\subsection{Spatial Distribution}  

\begin{figure*}[htb!]
    \centering
   %\epsscale{1}
%   \plotone
 \includegraphics[width=1\textwidth]{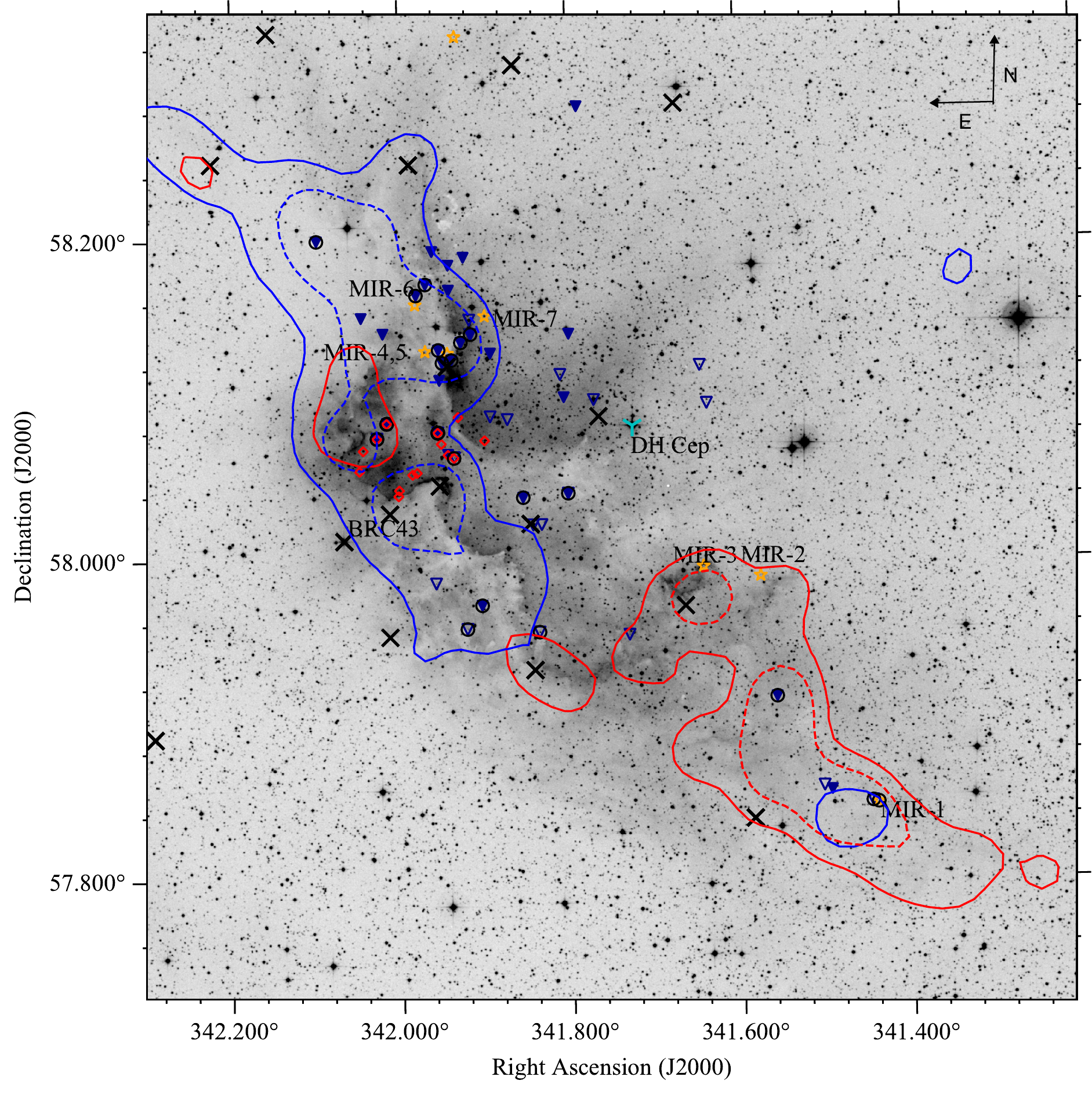}

           \caption{Location of all the young stellar objects, with the symbols the same as in Figure~\ref{fig:allccd}, on a DSS2 red image, together with IRAS point sources \citep{1988iras....7.....H} as protostar candidates (black crosses).  The Class~I (MIR) sources and DH\,Cep are labeled.  Overlaid are contours for emission from $^{12}$CO (solid contours) and $^{13}$CO (dashed contours) integrated in the velocity range of $-45$~km~s$^{-1}$ to $-39$~km~s$^{-1}$ in blue, and $-39$~km~s$^{-1}$ to $-35$~km~s$^{-1}$ in red. For clarity, only the outermost contours are shown at 5$\sigma$ level, with $\sigma$ being the sky background, to trace the cloud perimeter.}
    \label{fig:spatial_dist}
\end{figure*}

With the list of the young population at different evolutionary stages, we are ready to delineate their relative locations with respect to parental/remnant gas.  Figure~\ref{fig:spatial_dist} displays their positions, now including also those of IRAS point sources as potential protostars.  The young stars are numerous in the northeastern part of the cloud and are concentrated around BRC\,43. To the northeast, where the molecular gas interacts with the ionization front, there are one IRAS source and three Class~I (MIR-4, MIR-5, and MIR-6) sources, together with a young group of PMS objects associated with the cloud, some located near the rim.  Outside the cloud, there is only one Class~I source (MIR-7), and there are several T~Tauri stars, all aligned in the direction to DH\,Cep.  This MIR-7 source, earlier considered as a field variable star \citep{2016MNRAS.456.2505L}, in fact exhibits a Class~I SED, and has a proper motion ($-2.53$,$-2.12$)~mas~yr$^{-1}$ and a parallax of $0.37\pm0.04$~mas \citep{2020yCat.1350....0G}, consistent with being a member of NGC\,7380, which has an average proper motion of ($-2.51$,$-2.14$)~mas~yr$^{-1}$, and a parallax of $0.33\pm0.05$~mas \citep{2018A&A...618A..93C}.

In the BRC\,43 extension, five out of 14 H$\alpha$ objects (limited to this region) are common to the classical T~Tauri stars in our sample, whereas the remaining H$\alpha$ objects are possibly weak-lined T-Tauri stars.  All these line up roughly toward DH\,Cep as well.  Our analysis of the IR data shows no Class~I or Class~II objects inside BRC\,43, except a previously known IRAS point source \citep{1991ApJS...77...59S} and a sub-mm source \citep{2008A&A...477..557M}.  As in the northeast extension, here the youngest objects, those encircled in Figure~\ref{fig:final_cmd}, seem preferentially associated with dense parts of the cloud.    

In addition to these two extensions, there are less prominent concentrations revealed by the CO data, notably the one in the southwest that harbors an IRAS source and with MIR-2 and MIR-3 located near its periphery, tantalizingly enough, facing DH\,Cep. The southernmost reach of the complex, distant from the massive stars and from ionized gas, still shows the presence of a few PMS objects and a protostar (MIR-1). Unlike the northeastern lobe where young stars are present inside and outside of the remnant molecular clouds that must have been carved out, here there is a paucity of relatively evolved young stars, and existing young stars have all been formed in situ within the comparatively unperturbed cloud.

%fig12

%sec 5
\section{Discussion}\label{discussion}

The region of our study, comprising with a full-fledged star cluster, highly luminous stars photoionizing surrounding gas, and molecular cloud complex that must have partly given birth to these stars, yet partly survived the fierce environments to sustain present-day star formation, provides a plausible scenario of stellar feedback.  The relational sequence in position and age of next-generation young stars with respect to the parental clouds along with their menifestations as individual globules, or bright-rimmed clouds,  and to the massive stars serve as vital tools to identify prompted star formation. Such configuration is amply exhibited, e.g., in the Orion clouds \citep{1989ApJ...342L..87S,  2005ApJ...624..808L, 1995MNRAS.276..923R, 2008ApJ...687.1303G, 1992A&A...261..589C}.

\begin{figure*}[htbp!]
    \centering
    \includegraphics[width=0.8\textwidth]{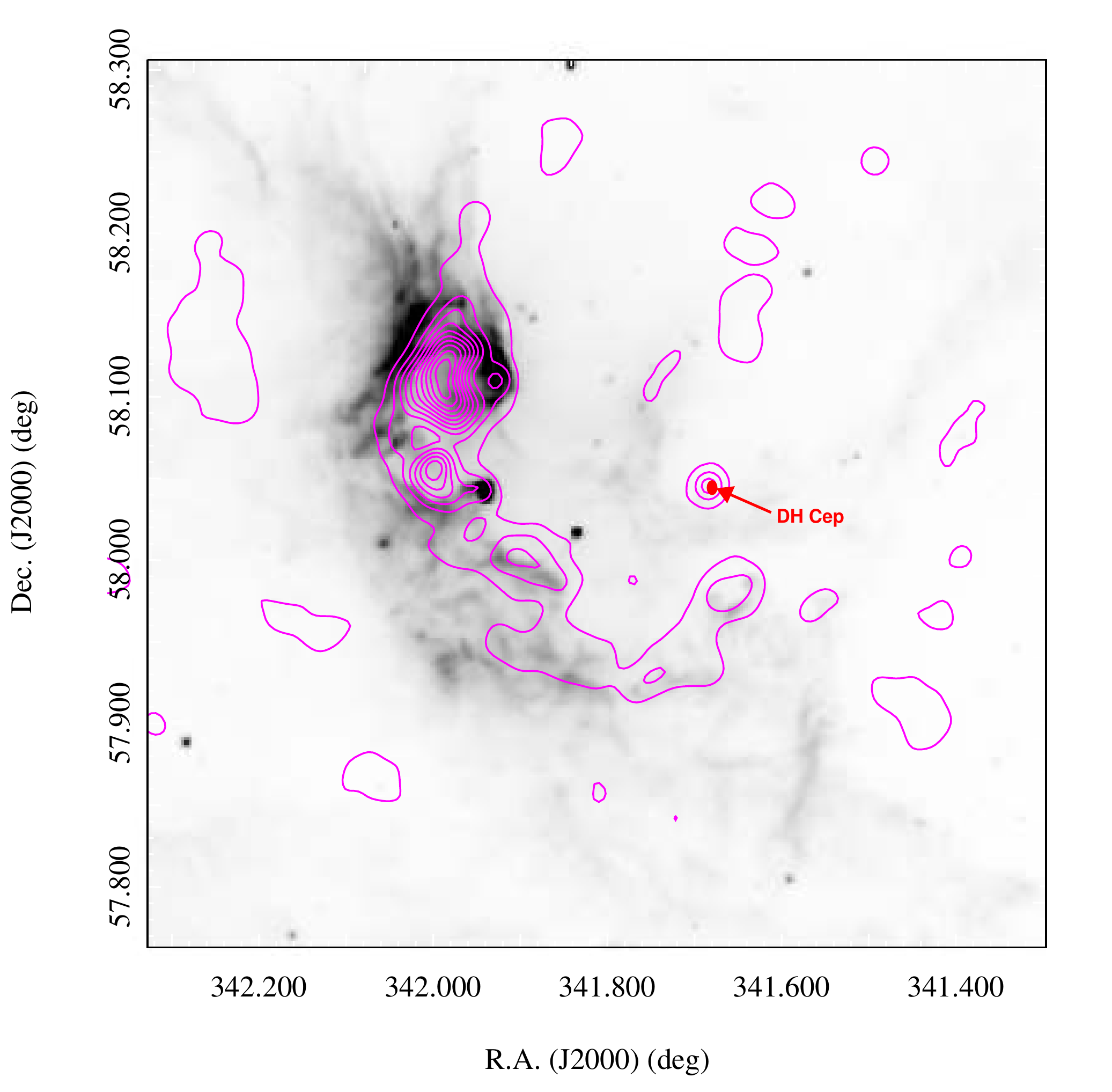}
    \caption{The NVSS 1.4~GHz continuum emission (contours in magenta) tracing hot plasma overlaid on the WISE 12~$\micron$ image (in grayscale). The lowest contour level is at 3$\sigma$, with $\sigma$ being 0.0005~Jy~beam$^{-1}$, with increments of 0.004~Jy~beam$^{-1}$.}
    \label{fig:1.4Ghz_w3}
\end{figure*}

Our analysis indicates that most Class~I and IRAS objects, signposts of ongoing star formation, are embedded and located near rims.  Moreover, the young stars line up roughly perpendicular to the rims, with bluer (presumably older) stars closer to, thereby exposing a starbirth sequence is therefore shown to start \citep{1995ApJ...455L..39S, 2002AJ....123.2597O, 2005ApJ...624..808L, 2007PASJ...59..199O, 2009MNRAS.396..964C}.  In addition to the distribution of molecular material, Figure~\ref{fig:1.4Ghz_w3} presents the associated ionized gas distinctly compressed in the northeastern extension, through BRC\,43, and stretching further to the upper southwestern lobe.  This is in support of an expanding ionization front originated from DH\,Cep.

%fig13

\begin{figure*}[htbp!]
    \centering
    \includegraphics[width=0.8\textwidth]{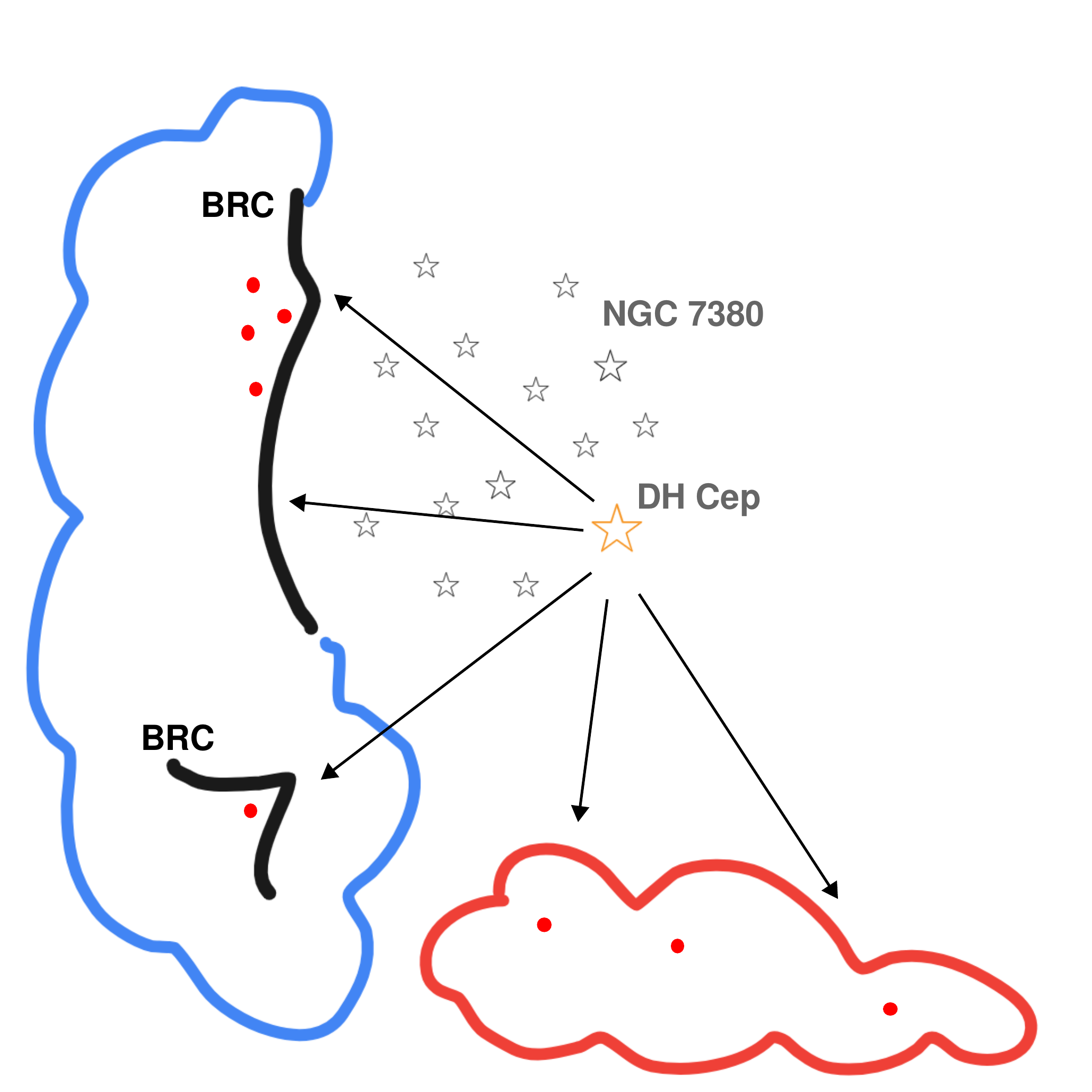}
    \caption{The schematic (not to scale) for the Sh2-142 region.  The asterisks represent the open cluster NGC\,7380, the orange asterisk the major exciting star DH~Cep.  Bright-rimmed clouds are marked, along with the red- and blue-shifted lobes of the molecular gas. The red dots show the youngest objects ($\la1$~Myr). The arrows emanating from DH\,Cep show possible propagation directions of ionizing fronts. }
    \label{fig:schematic_sh142}
\end{figure*}

The feasibility of a triggering scenario must be substantiated by reasonable time scales \citep{2003ApJ...595..900K,2005ApJ...624..808L}. To consider an extant OB star still in existence as the external triggerer, its age must be greater or comparable to the propagation time of the ionization front through the cloud, plus the age of the young population.  In our case, the expected main-sequence lifetime of DH\,Cep is $\sim1.5$~Myr \citep{1996A&A...314..165H}, and given a projected distance of 5.6~pc between BRC\,43 and DH\,Cep, the propagation time for the perturbing front, assuming a typical sound speed of $10$~km~s$^{-1}$, is $\sim 0.5$~Myr.  The northeastern extension is physically closer to, and hence reasonably coeval with, NGC\,7380.  Since the youngest population is $\la1$~Myr old, these timescales are consistent with a sequential star formation in the region. Figure \ref{fig:schematic_sh142} presents the schematic of the cloud complex comprising its main components namely, cluster, DH Cep, BRCs along with relatively  young population associated with the cloud, to sum up our findings and inference. 

There are limits and uncertainties of our interpretation.  Massive stars do not always play a constructive role to clouds that otherwise would not collapse spontaneously, as is the case proposed in this work. For example, they could disrupt a cloud so as to quench any consequent formation possibilities.  On the stellar scales, a young stellar disk/envelope can be photoevaporated by scorching radiation from nearby massive stars \citep{2009A&A...504...97B}, or swept by their winds, disguising a bona fide infant object as being devoid of circumstellar matter, hence seemingly evolved. This will affect the level of IR excess as amount of circumstellar dust which is a key indicator of its youth of an object, and its evolutionary stage. Therefore the age argument plus the non-random spatial distribution of the young population presented here lends support of decisive influence of massive stars on the environments in case of prompted star formation.

%fig14

\begin{acknowledgments}
TS and WPC acknowledge the financial support of this work by the MoST grant 109-2112-M-008-015-MY3. This research made use of the data from the Milky Way Imaging Scroll Painting (MWISP) project, which is a multi-line survey in $^{12}$CO/$^{13}$CO/C$^{18}$O along the northern galactic plane with PMO-13.7m telescope. We are grateful to all the members of the MWISP working group, particularly the staff members at PMO-13.7m telescope for their long-term support. MWISP was sponsored by National Key R\&D Program of China with grant 2017YFA0402701 and CAS Key Research Program of Frontier Sciences with grant QYZDJ-SSW-SLH047. YS is supported by National Natural Science Foundation of China through grant 11773077 and the Youth Innovation Promotion Association, CAS (2018355). Y.G. acknowledges support by National Key Basic R\&D  Program of China (2017YFA0402704), NSFC grant 11861131007, 12033004. This publication makes use of data products from the Two Micron All Sky Survey, which is a joint project of the University of Massachusetts and the Infrared Processing and Analysis Center/California Institute of Technology, funded by the National Aeronautics and Space Administration and the National Science Foundation. This publication makes use of data products from the Wide-field Infrared Survey Explorer, which is a joint project of the University of California, Los Angeles, and the Jet Propulsion Laboratory/California Institute of Technology, funded by the National Aeronautics and Space Administration. This work has made use of data from the European Space Agency (ESA) mission
{\it Gaia} (\url{https://www.cosmos.esa.int/gaia}), processed by the {\it Gaia}
Data Processing and Analysis Consortium (DPAC,
\url{https://www.cosmos.esa.int/web/gaia/dpac/consortium}). Funding for the DPAC
has been provided by national institutions, in particular the institutions
participating in the {\it Gaia} Multilateral Agreement.This research has made use of the services of the ESO Science Archive Facility
%This research has made use of the services of the ESO Science Archive Facility
\end{acknowledgments}

\appendix

\section{Catalog of Young Stars}

In addition to the Class~I objects listed in Table \ref{Tab:Table_sed}, we present in  Table~\ref{Tab:all_list} other classes of young stars, including H$\alpha$ emission stars, near-infrared and mid-infrared excess stars  identified from the 2MASS and WISE data respectively. Table~\ref{Tab:all_list} lists the $V$ magnitude, $J-H$ and $H-K_s$ colors, WISE colors and Gaia EDR3 measurements.

\begin{longrotatetable}
\begin{deluxetable}{LLLRRRRRRhRRl}

%% Keep a portrait orientation

%% Over-ride the default font size
%% Use 8pt
\tabletypesize{\scriptsize}

\tablecaption{List of Young Stars in Sh\,2-142 }

\tablenum{2}

\tablehead{
\colhead{ID} & \colhead{$\alpha$ ~ (J2000)} & \colhead{$\delta$ ~ (J2000)}  & \colhead{Gaia ID} & \colhead{$V$} & \colhead{$J-H$} & \colhead{$H-K_s$} & \colhead{$[3.4]-[4.6]$} & \colhead{$[12]-[22]$} & \colhead{} & \colhead{$\mu_{\alpha} \cos\delta$} & \colhead{$\mu_{\delta}$} & \colhead{Remarks \tablenotemark{$^\dagger$} } \\
\colhead{} & \colhead{ ~ (deg)} & \colhead{ ~ (deg)} & \colhead{} & \colhead{ ~ (mag)} & \colhead{ ~ (mag)} & \colhead{ ~ (mag)} & \colhead{ } & \colhead{ } & \colhead{} & \colhead{ ~ (mas~yr$^{-1}$)} & \colhead{ ~ (mas~yr$^{-1}$)} & \colhead{} 
         }
\startdata
1  &  341.900000  &  58.075556  &  2007418893273376384  &  16.777~(0.004) &  0.679  &  0.222  &  -0.124  &  0.275  &  0.1852~(0.1894) &  -2.1311~(0.192) &  -1.835~(0.204) &  H$\alpha$ \\
2  &  341.931250  &  58.090278  &  2007418996344878080  &  19.110~(0.011) &  0.948  &  0.406  &  \nodata &  \nodata &  -0.3401~(0.1100) &  -2.1633~(0.116) &  -2.396~(0.109) &  H$\alpha$, $a$  \\
3  &  341.936250  &  58.064722  &  2007418717164267776  &  19.414~(0.012) &  1.129  &  0.664  &  0.682  &  2.421  &  0.3485~(0.1293) &  -2.4842~(0.139) &  -1.578~(0.141) &  H$\alpha$ \\
4  &  341.943333  &  58.066667  &  2007418717170804224  &  20.997~(0.057) &  1.049  &  0.683  &  \nodata &  \nodata &  0.6829~(0.2371) &  -2.5157~(0.258) &  -2.399~(0.243) &  H$\alpha$ \\
5  &  341.951250  &  58.073611  &  2007418927625454208  &  20.188~(0.023) &  1.142  &  0.484  &  \nodata &  \nodata &  0.2012~(0.1973) &  -2.2366~(0.216) &  -1.482~(0.198) &  H$\alpha$, NIR \\
6  &  341.955417  &  58.080556  &  2007418927633127936  &  21.129~(0.055) &  1.439  &  1.066  &  0.655  &  2.718  &  -1.5337~(0.5589) &  0.6507~(0.600) &  -2.681~(0.547) &  H$\alpha$, NIR, TD, $a, b$  \\
7  &  341.980000  &  58.055556  &  \nodata &  19.878~(0.017) &  \nodata &  \nodata &  \nodata &  \nodata &  \nodata &  \nodata &  \nodata &  H$\alpha$ \\
8  &  341.985417  &  58.054444  &  2007371747402090752  &  19.855~(0.018) &  1.062  &  0.627  &  \nodata &  \nodata &  0.4861~(0.2089) &  -2.1438~(0.225) &  -1.588~(0.202) &  H$\alpha$ \\
9  &  342.001250  &  58.044722  &  2007371682977631744  &  18.837~(0.008) &  1.100  &  0.423  &  0.392  &  1.700  &  0.3197~(0.2832) &  1.5372~(0.293) &  -5.622~(0.270) &  H$\alpha$, $b$ \\
10  &  342.002083  &  58.041111  &  2007371678682606208  &  21.016~(0.044) &  0.793  &  0.612  &  \nodata &  \nodata &  -0.5921~(0.3748) &  -2.5072~(0.397) &  -3.253~(0.362) &  H$\alpha$, $a$ \\
11  &  342.015833  &  58.086389  &  2007420160273312512  &  20.129~(0.020) &  1.590  &  1.172  &  0.578  &  1.475  &  2.4237~(0.2641) &  -1.7524~(0.267) &  -2.218~(0.245) &  H$\alpha$ , NIR, $a$\\
12  &  342.027500  &  58.076944  &  2007373263538670976  &  18.389~(0.006) &  1.342  &  0.985  &  0.642  &  2.868  &  0.3252~(0.0563) &  -2.0504~(0.058) &  -2.175~(0.057) &  H$\alpha$, NIR, MIR \\
13  &  342.043750  &  58.069444  &  2007373194819202944  &  17.635~(0.005) &  0.980  &  0.383  &  0.119  &  2.791  &  0.3701~(0.3625) &  -1.0993~(0.365) &  -3.017~(0.361) &  H$\alpha$, $b$ \\
14  &  342.048333  &  58.056667  &  2007372984352672896  &  18.228~(0.007) &  0.609  &  0.002  &  \nodata &  \nodata &  0.4374~(0.0824) &  -4.5821~(0.086) &  -3.206~(0.079) &  H$\alpha$, $b$ \\
15  &  341.494397  &  57.855476  &  2007409783635818240  &  \nodata &  0.952  &  0.627  &  0.271  &  3.879  &  0.2884~(0.1585) &  -2.7482~(0.170) &  -1.939~(0.156) &  NIR \\
16  &  341.557581  &  57.914074  &  2007410230308741376  &  \nodata &  0.962  &  0.591  &  0.452  &  2.100  &  0.3885~(0.0820) &  -2.6836~(0.089) &  -2.028~(0.087) &  NIR \\
17  &  341.904388  &  57.972439  &  2007370854050001664  &  \nodata &  1.139  &  0.718  &  0.546  &  2.275  &  0.0417~(0.2449) &  -2.5860~(0.269) &  -2.610~(0.242) &  NIR \\
18  &  341.801554  &  58.042183  &  2007419099424074752  &  \nodata &  1.099  &  0.703  &  0.698  &  2.815  &  0.8689~(0.4292) &  -0.9822~(0.469) &  -4.302~(0.490) &  NIR, MIR, $b$ \\
19  &  341.854893  &  58.039734  &  2007418412237051008  &  18.427~(0.009) &  1.029  &  0.629  &  0.419  &  1.548  &  1.1627~(0.3412) &  -3.1667~(0.346) &  0.193~(0.355) &  NIR, $b$ \\
20  &  341.805658  &  58.102238  &  2007422496742049664  &  \nodata &  1.116  &  0.727  &  0.609  &  1.433  &  1.3603~(0.3307) &  -3.0904~(0.349) &  -1.835~(0.297) &  NIR, $a$  \\
21  &  341.949521  &  58.124252  &  2007419855346026752  &  16.328~(0.003) &  0.889  &  0.560  &  0.503  &  4.240  &  0.3309~(0.0247) &  -2.3453~(0.026) &  -2.334~(0.025) &  NIR, TD \\
22  &  341.953041  &  58.113277  &  2007419782322726912  &  \nodata &  1.178  &  0.712  &  \nodata &  \nodata &  0.4052~(0.3351) &  -2.4728~(0.386) &  -1.795~(0.323) &  NIR \\
23  &  341.799320  &  58.141865  &  2007422879004712960  &  \nodata &  0.917  &  0.568  &  \nodata &  \nodata &  -0.7669~(0.8914) &  -2.6417~(0.980) &  0.845~(0.923) &  NIR, $a, b$   \\
24  &  341.915786  &  58.142029  &  2007420095864167552  &  17.847~(0.006) &  1.144  &  0.779  &  0.528  &  1.973  &  0.4839~(0.0732) &  -2.6501~(0.089) &  -2.103~(0.068) &  NIR, MIR\\
25  &  341.927422  &  58.136948  &  \nodata &  \nodata &  1.256  &  0.764  &  \nodata &  \nodata &  \nodata &  \nodata &  \nodata &  NIR \\
26  &  341.892512  &  58.130032  &  2007419992777147648  &  20.216~(0.034) &  0.954  &  0.673  &  \nodata &  \nodata &  -0.1131~(0.4632) &  -4.9446~(0.499) &  -0.178~(0.469) &  NIR, $a,b$ \\
27  &  341.968923  &  58.173332  &  2007424562622310400  &  \nodata &  1.248  &  0.754  &  -0.062  &  2.459  &  0.6113~(0.2318) &  -2.1720~(0.267) &  -2.457~(0.223) &  NIR \\
28  &  341.953892  &  58.132324  &  \nodata &  \nodata &  1.654  &  1.014  &  \nodata &  \nodata &  \nodata &  \nodata &  \nodata &  NIR \\
29  &  341.941872  &  58.185230  &  2007424627039547904  &  \nodata &  0.955  &  0.621  &  0.480  &  3.446  &  0.0829~(0.3162) &  -2.7824~(0.346) &  -2.440~(0.306) &  NIR \\
30  &  341.938608  &  58.125923  &  2007419851042201600  &  \nodata &  1.429  &  0.985  &  0.574  &  4.879  &  0.3389~(0.2389) &  -2.4699~(0.277) &  -2.460~(0.243) &  NIR, TD \\
31  &  341.941590  &  58.169617  &  2007423080851313664  &  \nodata &  1.065  &  0.658  &  0.092  &  -0.296  &  0.1464~(0.1478) &  -2.7917~(0.174) &  -2.365~(0.154) &  NIR \\
32  &  341.980183  &  58.166515  &  2007421642052396672  &  17.222~(0.011) &  0.906  &  0.613  &  \nodata &  \nodata &  0.3380~(0.0480) &  -2.2764~(0.051) &  -2.175~(0.045) &  NIR \\
33  &  342.019767  &  58.142502  &  2007421294154343296  &  19.486~(0.014) &  0.974  &  0.767  &  \nodata &  \nodata &  0.6906~(0.1668) &  -3.7559~(0.179) &  -1.024~(0.160) &  NIR, $b$ \\
34  &  342.045459  &  58.152340  &  2007421397224336000  &  \nodata &  1.139  &  0.874  &  \nodata &  \nodata &  1.0695~(0.1405) &  -10.2152~(0.159) &  5.828~(0.138) &  NIR, $b$ \\
35  &  341.923714  &  58.190090  &  2007423355735513472  &  \nodata &  0.942  &  0.602  &  \nodata &  \nodata &  0.5070~(0.2245) &  -3.3429~(0.253) &  -2.072~(0.220) &  NIR \\
36  &  341.960885  &  58.193882  &  2007424631341712512  &  \nodata &  1.174  &  0.847  &  \nodata &  \nodata &  0.5393~(0.2330) &  -2.3024~(0.268) &  -1.781~(0.224) &  NIR \\
37  &  342.097888  &  58.200584  &  2007422123088832000  &  16.610~(0.012) &  0.972  &  0.596  &  -0.003  &  3.270  &  0.7183~(0.0377) &  -11.6709~(0.041) &  -3.541~(0.037) &  NIR, $b$ \\
38  &  341.787287  &  58.284023  &  2007437305785335168  &  \nodata &  0.875  &  0.556  &  0.274  &  0.767  &  0.4866~(0.1866) &  -2.7023~(0.187) &  -2.635~(0.170) &  NIR \\
39  &  342.113152  &  58.353428  &  \nodata &  \nodata &  1.112  &  0.741  &  0.194  &  2.655  &  \nodata &  \nodata &  \nodata &  NIR \\
40  &  341.922024  &  57.957687  &  2007370102442790400  &  \nodata &  1.255  &  0.714  &  0.934  &  2.761  &  0.1644~(0.1255) &  -2.5693~(0.132) &  -2.603~(0.129) &  MIR \\
41  &  341.958527  &  57.986312  &  2007370510452682624  &  19.571~(0.013) &  1.124  &  0.812  &  0.651  &  2.340  &  0.4515~(0.1123) &  -2.3725~(0.121) &  -2.539~(0.117) &  MIR, TD \\
42  &  341.836899  &  57.955555  &  2007370785329140480  &  \nodata &  1.176  &  0.636  &  0.540  &  2.083  &  0.1698~(0.1653) &  -2.2811~(0.179) &  -0.623~(0.164) &  MIR \\
43  &  342.480315  &  58.197335  &  2007398444923197184  &  \nodata &  0.498  &  0.796  &  0.859  &  2.815  &  -0.2364~(0.3566) &  0.0199~(0.403) &  0.056~(0.342) &  MIR, $a, b$ \\
44  &  341.446427  &  57.848265  &  2007409616143542144  &  \nodata &  0.952  &  0.452  &  0.951  &  2.444  &  -0.0550~(0.1448) &  -2.7023~(0.158) &  -3.459~(0.145) &  MIR, $a$ \\
45  &  341.843310  &  58.022632  &  2007418171718891904  &  14.878~(0.007) &  1.034  &  0.916  &  0.996  &  2.446  &  -0.8131~(0.1905) &  -1.7739~(0.191) &  -3.383~(0.207) &  MIR, $a$  \\
46  &  341.833242  &  58.022783  &  \nodata &  \nodata &  \nodata &  \nodata &  0.899  &  2.843  &  \nodata &  \nodata &  \nodata &  MIR \\
47  &  341.770388  &  58.100528  &  2007422462376242816  &  \nodata &  0.130  &  0.186  &  0.523  &  2.490  &  0.9336~(0.1975) &  -3.4756~(0.236) &  -0.342~(0.210) &  MIR, TD, $b$ \\
48  &  341.809836  &  58.116498  &  2007422604125045632  &  15.465~(0.004) &  0.710  &  0.799  &  0.708  &  3.409  &  0.4231~(0.0258) &  -2.5984~(0.028) &  -2.069~(0.025) &  MIR, TD  \\
49  &  341.917174  &  58.151031  &  2007423050801656960  &  13.544~(0.005) &  0.341  &  0.288  &  0.274  &  2.178  &  0.2133~(0.0124) &  -3.1701~(0.013) &  -1.777~(0.012) &  MIR, TD  \\
50  &  341.893151  &  58.090457  &  2007419644877416192  &  18.938~(0.010) &  0.830  &  0.494  &  0.488  &  2.755  &  0.2933~(0.1346) &  -2.3141~(0.146) &  -2.162~(0.134) &  MIR \\
51  &  341.503446  &  57.858029  &  2007409817998522368  &  \nodata &  1.331  &  1.178  &  0.709  &  2.590  &  \nodata &  \nodata &  \nodata &  MIR \\
52  &  341.636622  &  58.097843  &  2007417514575231232  &  \nodata &  0.860  &  0.459  &  0.675  &  2.768  &  0.5286~(0.2597) &  -2.6819~(0.313) &  -1.398~(0.289) &  MIR \\
53  &  341.730776  &  57.953564  &  2007411952606264448  &  \nodata &  1.034  &  0.532  &  0.533  &  1.935  &  \nodata &  \nodata &  \nodata &  MIR \\
54  &  341.872801  &  58.088600  &  \nodata &  \nodata &  \nodata &  \nodata &  0.735  &  3.036  &  \nodata &  \nodata &  \nodata &  MIR \\
\enddata
\label{Tab:all_list}

\end{deluxetable}

\tablenotetext{$$^\dagger$$}{H$\alpha$: H$\alpha$ emission line; NIR: near-infrared excess; MIR: mid-infrared excess; TD: transition disks}
\tablenotetext{$a$}{Parallax ($\varpi$) negative values or high values}
\tablenotetext{$b$}{Proper motion suggesting non-membership}

\end{longrotatetable}


\begin{thebibliography}{}
\expandafter\ifx\csname natexlab\endcsname\relax\def\natexlab#1{#1}\fi
\providecommand{\url}[1]{\href{#1}{#1}}
\providecommand{\dodoi}[1]{doi:~\href{http://doi.org/#1}{\nolinkurl{#1}}}
\providecommand{\doeprint}[1]{\href{http://ascl.net/#1}{\nolinkurl{http://ascl.net/#1}}}
\providecommand{\doarXiv}[1]{\href{https://arxiv.org/abs/#1}{\nolinkurl{https://arxiv.org/abs/#1}}}

\bibitem[{198(1988)}]{1988iras....7.....H}
 1988, {Infrared Astronomical Satellite (IRAS) Catalogs and Atlases.Volume 7:
  The Small Scale Structure Catalog.}, Vol.~7

\bibitem[{{Baade}(1983)}]{1983A&AS...51..235B}
{Baade}, D. 1983, \aaps, 51, 235

\bibitem[{{Bailer-Jones} {et~al.}(2021){Bailer-Jones}, {Rybizki}, {Fouesneau},
  {Demleitner}, \& {Andrae}}]{2021AJ....161..147B}
{Bailer-Jones}, C.~A.~L., {Rybizki}, J., {Fouesneau}, M., {Demleitner}, M., \&
  {Andrae}, R. 2021, \aj, 161, 147, \dodoi{10.3847/1538-3881/abd806}

\bibitem[{{Beltr{\'a}n} {et~al.}(2009){Beltr{\'a}n}, {Massi}, {L{\'o}pez},
  {Girart}, \& {Estalella}}]{2009A&A...504...97B}
{Beltr{\'a}n}, M.~T., {Massi}, F., {L{\'o}pez}, R., {Girart}, J.~M., \&
  {Estalella}, R. 2009, \aap, 504, 97, \dodoi{10.1051/0004-6361/200811540}

\bibitem[{{Bertoldi}(1989)}]{1989ApJ...346..735B}
{Bertoldi}, F. 1989, \apj, 346, 735, \dodoi{10.1086/168055}

\bibitem[{{Bessell} \& {Brett}(1988)}]{1988PASP..100.1134B}
{Bessell}, M.~S., \& {Brett}, J.~M. 1988, \pasp, 100, 1134,
  \dodoi{10.1086/132281}

\bibitem[{{Bhatt} {et~al.}(2010){Bhatt}, {Pandey}, {Kumar}, {Sagar}, \&
  {Singh}}]{2010NewA...15..755B}
{Bhatt}, H., {Pandey}, J.~C., {Kumar}, B., {Sagar}, R., \& {Singh}, K.~P. 2010,
  \na, 15, 755, \dodoi{10.1016/j.newast.2010.04.001}

\bibitem[{{Cantat-Gaudin} {et~al.}(2018){Cantat-Gaudin}, {Jordi}, {Vallenari},
  {Bragaglia}, {Balaguer-N{\'u}{\~n}ez}, {Soubiran}, {Bossini}, {Moitinho},
  {Castro-Ginard}, {Krone-Martins}, {Casamiquela}, {Sordo}, \&
  {Carrera}}]{2018A&A...618A..93C}
{Cantat-Gaudin}, T., {Jordi}, C., {Vallenari}, A., {et~al.} 2018, \aap, 618,
  A93, \dodoi{10.1051/0004-6361/201833476}

\bibitem[{{Carpenter}(2001)}]{2001AJ....121.2851C}
{Carpenter}, J.~M. 2001, \aj, 121, 2851, \dodoi{10.1086/320383}

\bibitem[{{Cernicharo} {et~al.}(1992){Cernicharo}, {Bachiller}, {Duvert},
  {Gonzalez-Alfonso}, \& {Gomez-Gonzalez}}]{1992A&A...261..589C}
{Cernicharo}, J., {Bachiller}, R., {Duvert}, G., {Gonzalez-Alfonso}, E., \&
  {Gomez-Gonzalez}, J. 1992, \aap, 261, 589

\bibitem[{{Chauhan} {et~al.}(2009){Chauhan}, {Pandey}, {Ogura}, {Ojha},
  {Bhatt}, {Ghosh}, \& {Rawat}}]{2009MNRAS.396..964C}
{Chauhan}, N., {Pandey}, A.~K., {Ogura}, K., {et~al.} 2009, \mnras, 396, 964,
  \dodoi{10.1111/j.1365-2966.2009.14756.x}

\bibitem[{{Chavarria-K} {et~al.}(1994){Chavarria-K}, {Moreno-Corral},
  {Hernandez-Toledo}, {Terranegra}, \& {de Lara}}]{1994A&A...283..963C}
{Chavarria-K}, C., {Moreno-Corral}, M.~A., {Hernandez-Toledo}, H.,
  {Terranegra}, L., \& {de Lara}, E. 1994, \aap, 283, 963

\bibitem[{{Chen} {et~al.}(2007){Chen}, {Lee}, \&
  {Sanchawala}}]{2007IAUS..237..278C}
{Chen}, W.~P., {Lee}, H.~T., \& {Sanchawala}, K. 2007, in IAU Symposium, Vol.
  237, Triggered Star Formation in a Turbulent ISM, ed. B.~G. {Elmegreen} \&
  J.~{Palous}, 278--282, \dodoi{10.1017/S1743921307001603}

\bibitem[{{Chen} {et~al.}(2011){Chen}, {Pandey}, {Sharma}, {Chen}, {Chen},
  {Sperauskas}, {Ogura}, {Chuang}, \& {Boyle}}]{2011AJ....142...71C}
{Chen}, W.~P., {Pandey}, A.~K., {Sharma}, S., {et~al.} 2011, \aj, 142, 71,
  \dodoi{10.1088/0004-6256/142/3/71}

\bibitem[{{Cohen} {et~al.}(1981){Cohen}, {Frogel}, {Persson}, \&
  {Elias}}]{1981ApJ...249..481C}
{Cohen}, J.~G., {Frogel}, J.~A., {Persson}, S.~E., \& {Elias}, J.~H. 1981,
  \apj, 249, 481, \dodoi{10.1086/159308}

\bibitem[{{Conrad} {et~al.}(2017){Conrad}, {Scholz}, {Kharchenko}, {Piskunov},
  {R{\"o}ser}, {Schilbach}, {de Jong}, {Schnurr}, {Steinmetz}, {Grebel},
  {Zwitter}, {Bienaym{\'e}}, {Bland -Hawthorn}, {Gibson}, {Gilmore},
  {Kordopatis}, {Kunder}, {Navarro}, {Parker}, {Reid}, {Seabroke}, {Siviero},
  {Watson}, \& {Wyse}}]{2017A&A...600A.106C}
{Conrad}, C., {Scholz}, R.~D., {Kharchenko}, N.~V., {et~al.} 2017, \aap, 600,
  A106, \dodoi{10.1051/0004-6361/201630012}

\bibitem[{{Cutri} \& {et al.}(2013)}]{2013yCat.2328....0C}
{Cutri}, R.~M., \& {et al.} 2013, VizieR Online Data Catalog, II/328

\bibitem[{{Cutri} {et~al.}(2003){Cutri}, {Skrutskie}, {van Dyk}, {Beichman},
  {Carpenter}, {Chester}, {Cambresy}, {Evans}, {Fowler}, {Gizis}, {Howard},
  {Huchra}, {Jarrett}, {Kopan}, {Kirkpatrick}, {Light}, {Marsh}, {McCallon},
  {Schneider}, {Stiening}, {Sykes}, {Weinberg}, {Wheaton}, {Wheelock}, \&
  {Zacarias}}]{2003yCat.2246....0C}
{Cutri}, R.~M., {Skrutskie}, M.~F., {van Dyk}, S., {et~al.} 2003, VizieR Online
  Data Catalog, II/246

\bibitem[{{Elmegreen}(1998)}]{1998ASPC..148..150E}
{Elmegreen}, B.~G. 1998, in Astronomical Society of the Pacific Conference
  Series, Vol. 148, Origins, ed. C.~E. {Woodward}, J.~M. {Shull}, \&
  J.~{Thronson}, Harley~A., 150.
\newblock \doarXiv{astro-ph/9712352}

\bibitem[{{Elmegreen} \& {Lada}(1977)}]{1977ApJ...214..725E}
{Elmegreen}, B.~G., \& {Lada}, C.~J. 1977, \apj, 214, 725,
  \dodoi{10.1086/155302}

\bibitem[{{Gaia Collaboration}(2020)}]{2020yCat.1350....0G}
{Gaia Collaboration}. 2020, VizieR Online Data Catalog, I/350

\bibitem[{{Gandolfi} {et~al.}(2008){Gandolfi}, {Alcal{\'a}}, {Leccia},
  {Frasca}, {Spezzi}, {Covino}, {Testi}, {Marilli}, \&
  {Kainulainen}}]{2008ApJ...687.1303G}
{Gandolfi}, D., {Alcal{\'a}}, J.~M., {Leccia}, S., {et~al.} 2008, \apj, 687,
  1303, \dodoi{10.1086/591729}

\bibitem[{{Georgelin} \& {Georgelin}(1976)}]{1976A&A....49...57G}
{Georgelin}, Y.~M., \& {Georgelin}, Y.~P. 1976, \aap, 49, 57

\bibitem[{{Hilditch} {et~al.}(1996){Hilditch}, {Harries}, \&
  {Bell}}]{1996A&A...314..165H}
{Hilditch}, R.~W., {Harries}, T.~J., \& {Bell}, S.~A. 1996, \aap, 314, 165

\bibitem[{{Israel}(1978)}]{1978A&A....70..769I}
{Israel}, F.~P. 1978, \aap, 70, 769

\bibitem[{{Joncas} {et~al.}(1985){Joncas}, {Dewdney}, {Higgs}, \&
  {Roy}}]{1985ApJ...298..596J}
{Joncas}, G., {Dewdney}, P.~E., {Higgs}, L.~A., \& {Roy}, J.~R. 1985, \apj,
  298, 596, \dodoi{10.1086/163644}

\bibitem[{{Joncas} {et~al.}(1988){Joncas}, {Koempe}, \& {de La
  Noe}}]{1988ApJ...332.1030J}
{Joncas}, G., {Koempe}, C., \& {de La Noe}, J. 1988, \apj, 332, 1030,
  \dodoi{10.1086/166710}

\bibitem[{{Joncas} \& {Roy}(1984)}]{1984ApJ...283..640J}
{Joncas}, G., \& {Roy}, J.~R. 1984, \apj, 283, 640, \dodoi{10.1086/162349}

\bibitem[{{Karr} \& {Martin}(2003)}]{2003ApJ...595..900K}
{Karr}, J.~L., \& {Martin}, P.~G. 2003, \apj, 595, 900, \dodoi{10.1086/376590}

\bibitem[{{Kharchenko} {et~al.}(2005){Kharchenko}, {Piskunov}, {R{\"o}ser},
  {Schilbach}, \& {Scholz}}]{2005A&A...438.1163K}
{Kharchenko}, N.~V., {Piskunov}, A.~E., {R{\"o}ser}, S., {Schilbach}, E., \&
  {Scholz}, R.~D. 2005, \aap, 438, 1163, \dodoi{10.1051/0004-6361:20042523}

\bibitem[{{Koenig} \& {Leisawitz}(2014)}]{2014ApJ...791..131K}
{Koenig}, X.~P., \& {Leisawitz}, D.~T. 2014, \apj, 791, 131,
  \dodoi{10.1088/0004-637X/791/2/131}

\bibitem[{{Koenig} {et~al.}(2012){Koenig}, {Leisawitz}, {Benford}, {Rebull},
  {Padgett}, \& {Assef}}]{2012ApJ...744..130K}
{Koenig}, X.~P., {Leisawitz}, D.~T., {Benford}, D.~J., {et~al.} 2012, \apj,
  744, 130, \dodoi{10.1088/0004-637X/744/2/130}

\bibitem[{{Lata} {et~al.}(2016){Lata}, {Pandey}, {Panwar}, {Chen}, {Samal}, \&
  {Pandey}}]{2016MNRAS.456.2505L}
{Lata}, S., {Pandey}, A.~K., {Panwar}, N., {et~al.} 2016, \mnras, 456, 2505,
  \dodoi{10.1093/mnras/stv2800}

\bibitem[{{Lee} \& {Chen}(2007)}]{2007ApJ...657..884L}
{Lee}, H.-T., \& {Chen}, W.~P. 2007, \apj, 657, 884, \dodoi{10.1086/510893}

\bibitem[{{Lee} {et~al.}(2005){Lee}, {Chen}, {Zhang}, \&
  {Hu}}]{2005ApJ...624..808L}
{Lee}, H.-T., {Chen}, W.~P., {Zhang}, Z.-W., \& {Hu}, J.-Y. 2005, \apj, 624,
  808, \dodoi{10.1086/429122}

\bibitem[{{Leisawitz} {et~al.}(1989){Leisawitz}, {Bash}, \&
  {Thaddeus}}]{1989ApJS...70..731L}
{Leisawitz}, D., {Bash}, F.~N., \& {Thaddeus}, P. 1989, \apjs, 70, 731,
  \dodoi{10.1086/191357}

\bibitem[{{Levreault}(1988)}]{1988ApJS...67..283L}
{Levreault}, R.~M. 1988, \apjs, 67, 283, \dodoi{10.1086/191275}

\bibitem[{{Meyer} {et~al.}(1997){Meyer}, {Calvet}, \&
  {Hillenbrand}}]{1997AJ....114..288M}
{Meyer}, M.~R., {Calvet}, N., \& {Hillenbrand}, L.~A. 1997, \aj, 114, 288,
  \dodoi{10.1086/118474}

\bibitem[{{Moffat}(1971)}]{1971A&A....13...30M}
{Moffat}, A.~F.~J. 1971, \aap, 13, 30

\bibitem[{{Morgan} {et~al.}(2010){Morgan}, {Figura}, {Urquhart}, \&
  {Thompson}}]{2010MNRAS.408..157M}
{Morgan}, L.~K., {Figura}, C.~C., {Urquhart}, J.~S., \& {Thompson}, M.~A. 2010,
  \mnras, 408, 157, \dodoi{10.1111/j.1365-2966.2010.17134.x}

\bibitem[{{Morgan} {et~al.}(2008){Morgan}, {Thompson}, {Urquhart}, \&
  {White}}]{2008A&A...477..557M}
{Morgan}, L.~K., {Thompson}, M.~A., {Urquhart}, J.~S., \& {White}, G.~J. 2008,
  \aap, 477, 557, \dodoi{10.1051/0004-6361:20078104}

\bibitem[{{Morgan} {et~al.}(2004){Morgan}, {Thompson}, {Urquhart}, {White}, \&
  {Miao}}]{2004A&A...426..535M}
{Morgan}, L.~K., {Thompson}, M.~A., {Urquhart}, J.~S., {White}, G.~J., \&
  {Miao}, J. 2004, \aap, 426, 535, \dodoi{10.1051/0004-6361:20040226}

\bibitem[{{Ogura} {et~al.}(2007){Ogura}, {Chauhan}, {Pandey}, {Bhatt}, {Ojha},
  \& {Itoh}}]{2007PASJ...59..199O}
{Ogura}, K., {Chauhan}, N., {Pandey}, A.~K., {et~al.} 2007, \pasj, 59, 199,
  \dodoi{10.1093/pasj/59.1.199}

\bibitem[{{Ogura} {et~al.}(2002){Ogura}, {Sugitani}, \&
  {Pickles}}]{2002AJ....123.2597O}
{Ogura}, K., {Sugitani}, K., \& {Pickles}, A. 2002, \aj, 123, 2597,
  \dodoi{10.1086/339976}

\bibitem[{{Pittard} \& {Stevens}(2002)}]{2002A&A...388L..20P}
{Pittard}, J.~M., \& {Stevens}, I.~R. 2002, \aap, 388, L20,
  \dodoi{10.1051/0004-6361:20020583}

\bibitem[{{Ramesh}(1995)}]{1995MNRAS.276..923R}
{Ramesh}, B. 1995, \mnras, 276, 923, \dodoi{10.1093/mnras/276.3.923}

\bibitem[{{Robitaille} {et~al.}(2007){Robitaille}, {Whitney}, {Indebetouw}, \&
  {Wood}}]{2007ApJS..169..328R}
{Robitaille}, T.~P., {Whitney}, B.~A., {Indebetouw}, R., \& {Wood}, K. 2007,
  \apjs, 169, 328, \dodoi{10.1086/512039}

\bibitem[{{Samal} {et~al.}(2012){Samal}, {Pandey}, {Ojha}, {Chauhan}, {Jose},
  \& {Pandey}}]{2012ApJ...755...20S}
{Samal}, M.~R., {Pandey}, A.~K., {Ojha}, D.~K., {et~al.} 2012, \apj, 755, 20,
  \dodoi{10.1088/0004-637X/755/1/20}

\bibitem[{{Schwartz}(1987)}]{1987ApJ...320..258S}
{Schwartz}, P.~R. 1987, \apj, 320, 258, \dodoi{10.1086/165537}

\bibitem[{{Sharpless}(1959)}]{1959ApJS....4..257S}
{Sharpless}, S. 1959, \apjs, 4, 257, \dodoi{10.1086/190049}

\bibitem[{{Siess} {et~al.}(2000){Siess}, {Dufour}, \&
  {Forestini}}]{2000A&A...358..593S}
{Siess}, L., {Dufour}, E., \& {Forestini}, M. 2000, \aap, 358, 593.
\newblock \doarXiv{astro-ph/0003477}

\bibitem[{{Su} {et~al.}(2019){Su}, {Yang}, {Zhang}, {Gong}, {Wang}, {Zhou},
  {Wang}, {Chen}, {Sun}, {Chen}, {Xu}, \& {Jiang}}]{2019ApJS..240....9S}
{Su}, Y., {Yang}, J., {Zhang}, S., {et~al.} 2019, \apjs, 240, 9,
  \dodoi{10.3847/1538-4365/aaf1c8}

\bibitem[{{Sugitani} {et~al.}(1989){Sugitani}, {Fukui}, {Mizuni}, \&
  {Ohashi}}]{1989ApJ...342L..87S}
{Sugitani}, K., {Fukui}, Y., {Mizuni}, A., \& {Ohashi}, N. 1989, \apjl, 342,
  L87, \dodoi{10.1086/185491}

\bibitem[{{Sugitani} {et~al.}(1991){Sugitani}, {Fukui}, \&
  {Ogura}}]{1991ApJS...77...59S}
{Sugitani}, K., {Fukui}, Y., \& {Ogura}, K. 1991, \apjs, 77, 59,
  \dodoi{10.1086/191597}

\bibitem[{{Sugitani} {et~al.}(1995){Sugitani}, {Tamura}, \&
  {Ogura}}]{1995ApJ...455L..39S}
{Sugitani}, K., {Tamura}, M., \& {Ogura}, K. 1995, \apjl, 455, L39,
  \dodoi{10.1086/309808}

\bibitem[{{Wilson}(1953)}]{1953GCRV..C......0W}
{Wilson}, R.~E. 1953, Carnegie Institute Washington D.C. Publication, 0

\end{thebibliography}
\end{document}